\documentclass[12pt,preprint]{aastex}

\shorttitle{Spatially Resolved H$_2$ from T Tauri Stars}
\shortauthors{Beck, McGregor \& Takami}

\begin{document}


\title{Spatially Resolved Molecular Hydrogen Emission in the Inner 200AU Environments of Classical T Tauri Stars\altaffilmark{1}}


\author{Tracy L. Beck\altaffilmark{2,3}, Peter J. McGregor\altaffilmark{4}, Michihiro Takami\altaffilmark{5,6} \& Tae-Soo Pyo\altaffilmark{5}}

\email{tbeck@stsci.edu, peter@mso.anu.edu.au, mtakami@subaru.naoj.org, pyo@subaru.naoj.org}

\altaffiltext{1}{Based on observations obtained at the Gemini Observatory, which is operated by the Association of Universities for Research in Astronomy, Inc., under a cooperative agreement with the NSF on behalf of the Gemini partnership: the National Science Foundation (United States), the Particle Physics and Astronomy Research Council (United Kingdom), the National Research Council (Canada), CONICYT (Chile), the Australian Research Council (Australia), CNPq (Brazil), and CONICET (Argentina).}
\altaffiltext{2}{Gemini North Observatory, 670 N. A'ohoku Pl. Hilo, HI 96720}
\altaffiltext{3}{Current Address: Space Telescope Science Institute, 3700 San Martin Dr., Baltimore, MD 21218}
\altaffiltext{4}{Research School of Astronomy \& Astrophysics,, Australian National University, Private Bag PO, Weston Creek, ACT 2611, Australia}
\altaffiltext{5}{Subaru Telescope, National Astronomical Observatory of Japan, 650 North  A`oh\=ok\=u Place, Hilo, HI 96720}
\altaffiltext{6}{Institute of Astronomy \& Astrophysics, Academia Sinica, P.O. Box 23-141, Taipei 10617, Taiwan, R. O. C.}



\begin{abstract}

We present 2.0-2.4$\mu$m integral field spectroscopy at adaptive optics spatial resolution ($\sim$0.''1) obtained with the Near-infrared Integral Field Spectrograph (NIFS) at Gemini North Observatory of six Classical T Tauri stars: T Tau, DG Tau, XZ Tau, HL Tau, RW Aur and HV Tau C.  In all cases, the {\it v}=1-0 S(1) (2.12$\mu$m) emission is detected at spatially extended distances from the central stars.  HL Tau, T Tau and HV Tau C have H$_2$ emission that extends to projected distances of more than $\sim$200 AU from the stars.  The bulk of the H$_2$ emission is typically not coincident with the location of continuum flux.  The observed morphologies vary between emission that is spatially continuous but decreasing away from the star (HV Tau C, DG Tau), and H$_2$ that shows discrete knots and arcs with 0.''2-0.''3 ($\sim$28-42 AU) spatial extents (T Tau, XZ Tau).  Multiple transitions detected in the K-band spectra show that H$_2$ level populations are typical of gas in thermal equilibrium with excitation temperatures in the 1800K-2300 K range.  Two dimensional maps of the extinction and H$_2$ excitation temperature were estimated from the high signal-to-noise (S/N) observations of T Tau using maps of the spatially resolved {\it v}=1-0 Q(3)/{\it v}=1-0 S(1) and {\it v}=2-1 S(1)/{\it v}=1-0 S(1) line ratios.  These maps show that A$_v$ and T$_{ex}$ values in the vicinity of T Tau vary on a significant level on spatial scales of less than 100 AU.  Three of the stars have H$_2$ velocity profiles that are centered at the stellar radial velocity, and three show velocity shifts with respect to the system.  RW Aur exhibits high and low velocity H$_2$ emission, and the observed morphology and kinematics of the high velocity material are consistent with atomic emission from its red-shifted micro-jet.  The morphologies detected in H$_2$ from DG Tau and HV Tau C resemble emission seen in scattered light from stars with inclined or edge-on circumstellar disks.  The locations and relative brightnesses of H$_2$ knots detected within $\sim$1$''$ of T Tau has varied on a timescale of $\sim$3 years.  Each of the stars studied here show observed excitation temperatures, spatial extents, and kinematics of the H$_2$ that are most consistent with shock excited emission from the inner regions of the known Herbig-Haro energy flows or from wide-angle winds encompassing the outflows rather than predominantly from UV or X-ray stimulated emission from the central stars.  The data presented in this study highlights the sensitivity of adaptive optics-fed integral field spectroscopy for spatially resolving emission line structures in the environments of bright young stars.

\end{abstract}




\section{Introduction}

The {\it v}=1-0 S(1) ro-vibrational transition of molecular hydrogen at 2.12$\mu$m was discovered in the spectra of a range of astronomical objects during the rapid expansion of infrared (IR) astronomy capabilities in the 1970s.  In star formation science, this emission was first detected from shock excited outflows in the Orion nebula (Gautier et al. 1976).  Since this time, molecular hydrogen from electronic transitions in the ultraviolet (UV), ro-vibrational transitions in the near-IR, and pure rotational emission in the mid and far IR have been studied in the spectra of young stars and their environs (Ardila et al. 2002; Saucedo et al. 2003; Walter et al. 2003; Herczeg et al. 2004; 2006; Davis et al. 2001; 2002; Bary et al. 2003; Takami et al. 2004; Duchene et al. 2005; Thi et al. 1999; 2001; Richter et al. 2002; Sheret et al. 2003; Sako et al. 2005).  Today, detection of {\it v}=1-0 S(1) H$_2$ (2.12$\mu$m) emission in the spectra of Classical T Tauri Stars (CTTSs), embedded proto-stars, and Herbig-Haro (HH) energy sources is commonplace.

Molecular hydrogen is a primary constituent of cool gas in the circumstellar environments of young stars.  A better understanding of the evolution of the gas, and of the H$_2$, is needed to ascertain how this material in circumstellar environments will ultimately evolve into planets.  Yet, many studies of molecular hydrogen emission in the inner $\sim$200 AU regions of CTTSs are for systems that are known to drive HH outflows (Saucedo et al. 2003; Walter et al. 2003; Herczeg et al. 2006; Takami et al. 2004; Kasper et al. 2002; Duchene et al. 2005).  In these inner environments of CTTSs, the main H$_2$ stimulation mechanisms, kinematics and morphological character from gas in outflows versus in quiescent circumstellar disks is not well established.  However, understanding the nature of the H$_2$ is especially of interest in the inner $\sim$50 AU where thermal emission from gas in the planet-forming regions of circumstellar disks is also believed arise (Nomura \& Millar 2005; Nomura et al. 2007; Bary et al. 2003; Ramsay Howat \& Greaves 2007).  

The ro-vibrational transitions of molecular hydrogen observed in the near IR spectra of young stars can be excited by UV fluorescence pumped by FUV or Ly $\alpha$ photons, stellar UV or X-ray heating, and by shocks in outflows.  Shock-excited H$_2$ typically shows thermal level populations in ro-vibrational features detected in K-band spectra (2.0-2.4$\mu$m), sometimes with high and low H$_2$ velocity features arising from different environments in the outflow or wind (Burton et al. 1989; Curiel et al. 1995; Eisloffel et al. 2000; Davis et al. 2001; 2002; Giannini et al. 2002; Takami et al. 2004; Takami et al. 2007).  In the case of UV fluorescence, molecular hydrogen is pumped into an electronically-excited state by Ly$\alpha$ or FUV flux (Ardila et al. 2002; Walter et al. 2003; Herczeg et al. 2004; 2006), and lines in the near IR arise from the corresponding cascade through the vibration-rotation transitions (Black \& van Dishoeck 1987; Sternberg \& Dalgarno 1989; Shull 1978).  Electronically excited states (and subsequently, high-v states) can be stimulated by UV fluorescence pumped by Ly$\alpha$ arising either from the star or in shocks, and in regions of low gas density the resulting level populations from this excitation can be quite different from thermal excitation (Black \& van Dishoeck 1987; Hasegawa et al. 1987; Ardila 2002).  Stellar UV and X-ray flux from the central star also heat the ambient circumstellar gas and dust.  The H$_2$ molecules are correspondingly excited into thermal equilibrium and level populations of low-v states can be identical to those observed in shocked regions (Maloney et al. 1996; Tine et al. 1997; Nomura \& Millar 2005; Nomura et al. 2007).  However, H$_2$ emission in the disk excited by UV pumping or heated by UV or X-ray photons from flux specifically arising from the young star is not expected to extend beyond a 30-50AU distance, and it should not be shifted in velocity by more than $\sim$10-15km/s (Maloney et al. 1996; Tine et al. 1997; Nomura \& Millar 2005; Nomura et al. 2007).

In CTTSs, {\it v}=1-0 S(1) ro-vibrational emission at 2.12$\mu$m and UV electronic transitions of molecular hydrogen were discovered in the spectra of the T Tau multiple system (Beckwith et al. 1978; Brown et al. 1981).  Since this time, T Tau has been a laboratory for studying the processes which govern molecular hydrogen in the inner environments of young stars.  IR and UV studies of T Tau have shown that extended knots and arcs of H$_2$ emission exist out to $\sim$20$''$ distances from the central stars (van Langevelde et al. 1994; Herbst et al. 1996; 1997; Walter et al. 2003).  These emission structures are best explained by shocks in extended regions of the known HH 155 and 255 outflows.  UV studies of the inner regions of the system have also found strong levels of fluorescent H$_2$ pumped by Ly$\alpha$ flux (Saucedo et al. 2003; Walter et al. 2003; Herczeg et al. 2006).  High spatial resolution imaging spectroscopy presented by Kasper et al. (2002) did not detect any near IR {\it v}=1-0 H$_2$ emission in the inner $\sim$70 AU region of T Tau, but the adaptive optics spectra of Duchene et al. (2005) did reveal significant emission near the position of the southern component, T Tau S.  Beck et al. (2004) also detected near-IR H$_2$ in emission toward T Tau S and found from multiple spectral measurements that that the line flux level could be variable on year-long timescales.  Herbst et al. (2007) has presented high spatial resolution imaging spectroscopy of H$_2$ in the region and found many spatially extended knots and arcs encompassing the position of T Tau S.  T Tau is arguably the best studied young star for understanding the processes that characterize molecular hydrogen in the inner circumstellar environments.  Yet, a clear definition of the location, structure and excitation of the H$_2$ has eluded astronomers because of different wavelength regions and sensitivities sampled by the observations and possible intrinsic variation in the emission. 

To date, little is known about the spatial structure, excitation, and variability of near IR molecular hydrogen emission in the inner $\sim$200 AU environments of most T Tauri stars.  In these regions, emission stimulated by UV pumped flux, UV and X-ray heating, and by shocks from known outflows are all expected to contribute to the observed H$_2$.  Investigation of the morphology, emission line ratios, and velocity structure in the H$_2$ transitions detectable in the K-band could distinguish between excitation mechanisms and clarify the nature of the emission in the inner regions of YSOs.  In just one pointing of a telescope, imaging spectroscopy with integral field units (IFUs) can provide three-dimensional x, y, $\lambda$ datacubes at high spatial resolution with simultaneous coverage of many molecular hydrogen emission lines of interest.  During the last few years, there has been a marked increase in the capabilities for adaptive optics (AO) fed near IR integral field spectroscopy at 8-10 meter class observatories (Eisenhauer et al. 2000; McGregor et al. 2003; Larkin et al. 2006).  IFUs optimized for AO spectroscopy have the power to spatially resolve molecular hydrogen emission structures with less than 0.$''$1 extents over the full wavelength ranges sampled by typical IR spectrographs.  As such, the new generation of IFUs provide the means to study the molecular hydrogen morphologies, excitations, and kinematics in the environments of CTTSs.  In this paper, we present the first star formation science results from the Near IR Integral Field Spectrograph at the Gemini North Observatory.  We report on the H$_2$ emission characteristics in the environments of six classic Classical T Tauri Stars:  T Tau, RW Aur, XZ Tau, DG Tau, HL Tau and HV Tau C. 

\section{Observations and Data Reduction}

Observations of the six CTTSs listed in Table 1 were obtained using the Near IR Intetral Field Spectrograph (NIFS) at the Frederick C. Gillette Gemini North Telescope on Mauna Kea, Hawaii.   NIFS is an image slicing IFU fed by Gemini's Near IR adaptive optics system, Altair, that is used to obtain integral field spectroscopy at spatial resolutions of $\le$0.$''$1 with a spectral resolving power of R$\sim$5300 at 2.2$\mu$m (as measured from arc and sky lines; McGregor et al. 2003).  The NIFS field has a spatial extent of 3$''\times3''$, and the individual IFU pixels are 0.$''$1$\times$0.$''$04 on the sky.  Data were obtained at the standard K-band wavelength setting for a spectral range of 2.003-2.447$\mu$m.  All observations were acquired in natural seeing of better than 0.$''$6 for excellent AO correction.  Unfortunately, thin cirrus was present for observations of HL Tau and T Tau, so absolute flux calibration was not done for the spectra presented here.

The data for this study were acquired for commissioning and system verification of NIFS in October 2005 and February 2006.  The six T Tauri stars that we observed were chosen for NIFS system tests because they are either bright enough to serve as their own adaptive optics wavefront reference stars, or they have nearby companions which served to test off-axis guide performance (e.g. HL Tau, HV Tau/C).  We also observed this sample of stars because they were known or suspected to have H$_2$ emission in their inner environments.  Table 1 summarizes the data obtained for this project.  Column 2 lists the HH object associated with each star and the third and fourth columns present the date our IFU spectra were acquired and the position angle of the observation.   In columns 5 and 6 the exposure time, coadds and number of individual exposures are listed.  Column 7 presents the total on-source exposure time for each star.  The last column in Table 1 notes the instrument system test that was done for each dataset.  Observations of DG Tau and T Tau were made for tests of the NIFS system flexure, and as a result very long on-source exposure times of 12120s and 4580s were acquired, respectively.  The NIFS On-Instrument Wavefront Sensor (OIWFS) provides slow flexure compensation in addition to the adaptive optics correction and was used for observations of DG Tau, XZ Tau and HL Tau.  For each observation, a standard set of calibrations were acquired using the Gemini facility calibration unit, GCAL.  The calibrations consisted of:  1) 5 flat fields taken with the IR continuum lamp on and 5 flats with lamps off, 2) an arc exposure using the Ar and Xe emission lamps and a corresponding dark taken with the same exposure time, and 3) a flat field frame taken through a Ronchi calibration mask which is used to rectify the spatial dimension in the IFU field.

The raw IFU frames were reduced into datacubes using the NIFS tasks in the Gemini IRAF package\footnote{Information on the Gemini IRAF package is available at http://www.gemini.edu/sciops/data/dataIRAFIndex.html}.  The {\bf nfprepare} task prepares the data for further reduction by attaching the Mask Definition File (MDF) to the image, adding a variance and data quality (DQ) plane, and verifying header keywords that define the instrument configuration.  The wrapper script {\bf nsreduce} then calls other tasks to cut the IFU data into 29 different extensions for each image slice in the science, variance and DQ planes (using {\bf nscut}), flat fields the data, and subtracts off a dark frame (where appropriate).  Data that were not acquired with an offset to a sky position required subtraction of the dark images to remove hot pixels from the NIFS detector. 

The data for RW Aur, HV Tau C and HL Tau were acquired by offsetting the telescope to a nearby blank sky field to measure the sky background brightness.  The data frames were reduced by subtracting the corresponding sky images from the science images.  For XZ Tau, T Tau and DG Tau, no sky offset images were obtained.  In these cases, a median sky spectrum was constructed by combining the data in the spatial dimension in 0.$''$15 circular apertures at the corners or edges of the final IFU datacube field.  The resulting sky emission spectrum was subtracted from each position in the science datacubes. The data for XZ Tau and DG Tau have very good sky subtraction because these datacubes have many spatial regions in where there is zero H$_2$ flux.  The field of T Tau has much more spatially extended H$_2$ emission, but we were able to construct a median sky spectrum from regions in the upper left corner and the middle right region of the field that had little or no H$_2$ emission.  The sky spectrum subtracted from T Tau had an H$_2$ flux of less than 3$\sigma$ level of detection above the noise, and we estimate that any excess H$_2$ emission subtraction for T Tau is at a level of less than 2-3\% of the flux in the regions that have H$_2$ detected at S/N=12.0 (correspondingly less at higher S/N).  We are confident that the H$_2$ emission from XZ Tau, DG Tau and T Tau is not severely affected by this method of sky subtraction, which worked very well in the absence of sky offset fields. 

After the images were processed through the {\bf nfreduce} task, they were run through {\bf nffixbad} to clean out residual bad pixels identified in the data quality planes.  The wavelength calibration and spatial rectification were applied to the data using the {\bf nffitcoords} and {\bf nftransform} routines; {\bf nffitcoords} determines the rectification mapping and adds calibration keywords to the image headers, and {\bf nftransform} applies the rectification to the data.  By inspecting the wavelength calibration and accuracy of sky emission lines positions, we estimate that the absolute velocity accuracy of our IFU data is $\sim$9-12 km/s (roughly one third of a spectral pixel).  The two dimensional images output by {\bf nftransform} were reconstructed into three dimensional datacubes (x, y and $\lambda$) using the {\bf nifcube} task.  For these datacubes, the spatial coordinates were remapped onto an interpolated scale with a square pixel grid and 0.$''$04/pixel extents on the sky.  Datacubes were generated for each of the on-source exposures we obtained.  The multiple files were merged into a single datacube for each target by determining a centroid position and shifting and coadding the individual datacubes.  This was accomplished using routines we have written to merge cubes in the IDL programming environment (Beck et al. 2004; 2007).  

We also observed A0 spectral type stars for removal of telluric absorption features with NIFS, and these data were processed into datacubes in the same manner described above.  These A0 stars were chosen based on their spectral types defined in the Hipparcos catalog and because they provided a good match to the airmass of the science target.  A pseudo longslit spectrum of each calibration star was extracted through the datacubes using a 1.$''$0 radius circular aperture using the IRAF task {\bf nfextract}.  These one dimensional spectra were divided by a 10,000K blackbody to remove the underlying shape of the stellar continuum.  A linear interpolation was done to remove the photospheric HI Br$\gamma$ absorption at 2.16$\mu$m in the A0 spectra.  The science datacubes of the YSOs were divided in the wavelength dimension by the spectra of the A0 calibrators using the task {\bf nftelluric}.  The long, on-source exposures for T Tau and DG Tau did not have interspersed telluric calibration observations, so correction of atmospheric features for these stars is poorer than might otherwise have been.  The imperfect atmospheric correction for T Tau and DG Tau is most problematic for the $\sim$2.35-2.44$\mu$m wavelength regions.  We estimate that the Q-branch emission fluxes are affected at a 10-15\% level for T Tau and $\sim$20\% for DG Tau because of this (Table 3).  The telluric correction in this wavelength region is particulary sensitive to variations and level of water vapor in the earths atmosphere.  However, the data for both T Tau and DG Tau were both acquired during extremely dry precipital water vapor conditions (less than 1.5mm of water vapor), so the effect of the imperfect telluric correction is much less than it might have been in wetter weather.  As discussed further in \S 3.1, observations for HL Tau were acquired in very wet weather, so the telluric correction as measured by the Q-branch flux levels is to be poor. The final reduced, combined and telluric corrected datacubes for each target are discussed in detail in the following sections.
 



\section{Results}

Adaptive optics fed near IR integral field spectroscopy provides information on molecular hydrogen emission line morphologies, excitation energies, and kinematics at projected spatial resolutions of $\sim$14 AU (0.$''$1) at the 140pc distance to the Taurus-Auriga dark clouds.  Figure 1 shows images of the continuum emission at 2.12$\mu$m for each of the six stars observed for this study.  These maps were constructed by fitting the continuum flux in the wavelength dimension and interpolating and removing the {\it v}=1-0 S(1) H$_2$ emission.  Residuals of this interpolation process were typically less than 1\% of the peak of the continuum emission.  Overplotted on the image of HL Tau is the position of the 0.$''$2 diameter occulting disk that was used for the observations to block the point source and increase the sensitivity to faint extended emission.  This occulting disk suppresses the light of HL Tau at a level of 10$^4$.  The average FWHM for the stars that appear as point sources is $\sim$0.$''$13 (DG Tau, T Tau, RW Aur and XZ Tau).   DG Tau is the only system which is a single star in these images.  HL Tau and HV Tau C have strong scattered light in their vicinity and are not point sources in the figure.  The edge-on disk morphology of HV Tau C is apparent in the continuum emission map (Monin \& Bouvier 2000; Stapelfeldt et al. 2003).  The multiple systems RW Aur AB, XZ Tau AB and T Tau North+South are easily separated in these data, but the 0.$''$1 separation T Tau South a/b system is not spatially resolved (Koresko 2000). 

\subsection{Spatially Extended H$_2$ Emission}

Figure 2 shows maps of the {\it v}=1-0 S(1) 2.12$\mu$m molecular hydrogen emission derived by subtracting the continuum maps from the IFU data and integrating over the wavelength extent of the H$_2$ feature.  Overplotted in blue are contours of the 2.12$\mu$m continuum emission from Figure 1.  Residuals of the subtraction process are sometimes noticeable at the positions of the continuum emission, but are typically less than 3\% of the peak flux.  Spatially extended H$_2$ emission is seen within 0.$''$2-1.$''$5 projected distances ($\sim$28-210 AU) of each of the six stars.  

Table 2 presents the fraction of the {\it v}=1-0 S(1) H$_2$ emission that is spatially coincident with the continuum K-band flux.  To derive these values, we constructed a median combined 1D spectrum at IFU spatial locations where the continuum flux was greater than 10\% of the peak value (a flux level of 5\% was used for T Tau to take into account the position of T Tau S).  The H$_2$ in this spectrum was then compared to a 1D spectrum of the H$_2$ emission at all other spatial locations in the IFU field, i.e., areas where the continuum flux was less than 10\% (5\% for T Tau).  The {\it v}=1-0 S(1) H$_2$ fluxes were determined by fitting and removing the continuum and integrating over the wavelength extent of the emission feature.  Typical uncertanties in the line fluxes were less than 5\%, and these were determined by comparing the peak emission flux to the range of the maxima and minima in the residuals of the continuum subtraction.  Over the spatial field, the percent of the H$_2$ flux that arose from the same spatial locations as the continuum emission is described by Table 2.  The long exposure observations of HL Tau were taken with the occulting disk, but an estimate of the on-source H$_2$ emission was made by scaling a short 60 second IFU spectrum acquired without the occulting disk for acquisition purposes.  We adopt a $\sim$15\% uncertainty in the on-source flux using this method, so the errors in Table 2 for HL Tau are correspondingly larger.  The majority of the detected H$_2$ 1-0 S(1) emission is not spatially coincident with the continuum flux for all stars except HV Tau C (Table 2).  This analysis suggests that the fraction of H$_2$ that is coincident with continuum emission is greatest for HV Tau C and HL Tau, the two stars that are viewed strongly in spatially resolved scattered light.

To further investigate the level of extended H$_2$, we also measured the emission line flux detected in spectra extracted with a 0.''2 circular apertures at the location of the peak continuum emission, and compared this to a spectrum extracted at the location of the peak of the spatially extended H$_2$ emission (Figure 2).  In all cases, H$_2$ emission was detected at the position of the central star.  From this analysis, RW Aur and XZ Tau represented the two extremes for low versus high H$_2$ flux at spatially extended locations.  The detected H$_2$ flux in the spatially extended spectrum was 27$\pm$5\% of the H$_2$ at the star for RW Aur, and for XZ Tau the extended H$_2$ had 420$\pm$20\% of the flux compared to the position of the brighter star.  If there existed a greater level of H$_2$ emission within 10-20 AU of the stars than for all other regions of the IFU field, the target H$_2$ fluxes should have been anywhere from $\sim$2 to 40-50 times greater than what we have detected at the on-source position. Based on this analysis and the data presented in Figure 2, we confidently claim that the bulk of H$_2$ emission in these CTTSs arises from regions in the IFU field that are spatially extended from the continuum emission.  We did not do this analysis for HV Tau C where the H$_2$ does not show strong peaks away from the continuum, and the on-source H$_2$ emission for HL Tau was estimated from the 60 second image obtained for acquisition.

Most of the H$_2$ emission from RW Aur and HL Tau is extended to one side of the continuum flux and is consistent with the locations of the inner jets associated with the HH 229 and 150 outflows, respectively (Pyo et al. 2006).  The H$_2$ emission from T Tau, HL Tau and XZ Tau shows several extended knots, while that from DG Tau and HV Tau C is more continuously distributed.   HV Tau C exhibits an edge-on disk in molecular hydrogen emission similar to the morphology seen in the continuum, but spatially extended wings of the H$_2$ appear to trace more flared regions of its circumstellar disk.  T Tau and HV Tau C both exhibit H$_2$ that extends over more than 60\% of the IFU field.  However, T Tau has a very complex morphology with both discrete knots and emission that is more spatially continuous.  The brightest area of H$_2$ emission from DG Tau shows a v-shaped morphology reminiscent of scattered light nebulosities associated with circumstellar disks viewed nearly edge-on (e. g. Padgett et al. 1999).  The H$_2$ emission from DG Tau has been attributed to a warm molecular wind encompassing the HH 158 outflow close to the star (Takami et al. 2004).  Discussion of the morphology of the H$_2$ and its origin in the environment of each star is continued in \S 4.1.

\subsection{H$_2$ Emission from Multiple Transitions: Line Ratios, Extinction and Excitation}

Figure 3 plots the K-band spectra for the six stars we observed for this study, and Table 3 shows the line fluxes for the H$_2$ features, normalized to the {\it v}=1-0 S(1) line.  Plotted in blue in each panel of Figure 3 is a spectrum of each YSO, centered at the peak of the continuum emission and extracted with the aperture size defined in the second column of Table 3 (upper region).  The H$_2$ fluxes in Table 3 were determined by fitting and removing the continuum and integrating over the wavelength extent of the emission feature.  We estimated uncertanties in the fluxes by comparing the peak emission flux to the range of the maxima and minima in the residuals of the continuum subtraction. This method resulted in conservative uncertainties for the line fluxes, These plots simulate the ``longslit'' K-band spectrum that would have been acquired acquired with a slit width corresponding to the extraction aperture size.  For RW Aur and HL Tau, the {\it v}=1-0 S(1) feature at 2.1218$\mu$m is the only H$_2$ line detected above the continuum flux in the spectra centered on the star.  Conversely, HV Tau C exhibits H$_2$ with high S/N from nine transitions in its K-band spectrum.

The H$_2$ emission is spatially extended from the continuum (Figure 2), and the strong stellar flux typically overwhelms the fainter H$_2$ lines for the spectra centered on the stars (for all but HV Tau C).  Hence, overplotted in green in Figures 3a-3e is a second spectrum that was extracted using a small, 0.$''$2 radius aperture centered on the peak of the {\it v}=1-0 S(1) emission line identified using the H$_2$ maps in Figure 2.  These spectra, plotted in green in Figure 3, demonstrate the increased sensitivity of the IFU to extended H$_2$ emission over traditional longslit spectroscopy (as approximated by the blue spectra).  H$_2$ Q-branch emissions (2.40-2.42$\mu$m) and higher vibrational level lines of {\it v}=2-1 S(1) (2.24$\mu$m) {\it v}=2-1 S(2) (2.15$\mu$m) and {\it v}=2-1 S(3) (2.07$\mu$m) are detected with a high level of significance in these spectra.  The corresponding line fluxes of these H$_2$ emission lines are included in the bottom region in Table 3.  The green spectra were extracted in smaller size apertures and show H$_2$ emission at a higher level of significance than the blue spectra.  Although both spectra are normalized to a value of 1.0 at 2.22$\mu$m in Figure 3, the true underlying continuum flux levels are vastly different in the two.  As a result of this and the fact that the apertures sizes are different, the data in Figure 3 do not represent the amount of H$_2$ emission in the on-source versus off-source spectra.  However, the total off-source flux in H$_2$ integrated over the full IFU spatial field is often greater than the on-source H$_2$, as described by Table 2.  

The spectra for HL Tau presented in Figure 3d have very different continuum shapes for the blue versus green plots.  This likely arises from different fractions of scattered versus transmissed continuum flux for HL Tau at different locations within the IFU field.  Takami et al. (2007) has shown evidence for scattered light from the outflow cavity walls near HL Tau, and if the H$_2$ emission is affected by scattering then it would make the continuum slope and line ratios appear bluer compared to reddening by ISM extinction.  The lower section in Table 3 also shows that the flux ratio of the Q(1) and Q(3) lines for HL Tau compared to Q(2) is only 1.0-1.9, whereas a wide range of physical excitation conditions can explain ratios in these lines of $\sim$2.5-4.0 (as seen for the other stars in our sample).  We have thus detected the Q(1) and Q(3) emisison in HL Tau at a 2.5-4 times weaker level than expected from typical H$_2$ excitation environments.  The most likely explanation for this discrepancy in Q-branch emission is, unfortunately, improper correction of the telluric features in the 2.40-2.44$\mu$m spectral region.   Particularly, the Q(1) and Q(3) branch H$_2$ emission features have telluric absorptions that lie at a similar wavelength, and accurate correction of these features is linked to the telluric H$_2$O vapor level.  Closer inspection of our data revealed that our observations of HL Tau in Feb. 2006 were acquired when the Q(1) and Q(3) emissions were shifted by heliocentric velocity effects to correspond precisely with the atmospheric absorption.  Additionally, the K-band NIFS spectroscopy for HL Tau were observed in high and variable water vapor conditions (precipital water vapor greater than 4.0mm).  Because of these reasons, extracting accurate fluxes in the Q-branch transitions was affected at significant levels by poor atmospheric correction in this wavelength region (estimated to be $>$50\% of the detected line flux). 

Molecular hydrogen transitions that arise from the same upper state have intrinsic flux ratios that are not affected by the excitation mechanism or physical characteristics of the emission environment.  As a result, the observed line ratios of the {\it v}=1-0 Q(3)/{\it v}=1-0 S(1) and {\it v}=1-0 Q(2)/{\it v}=1-0 S(0) transitions can provide a measure of the flux attenuation by extinction or scattering along the line of sight to the emitting region.  The intrinsic line ratio of the Q(3)/S(1) emission is 0.7 and the Q(2)/S(0) emission is 1.10 (Turner, Kirby-Docken \& Dalgarno 1977).  We assume that most of the attenuation in line emission comes from extinction, and derive values for the A$_v$ by dereddening the observed line fluxes assuming an ISM extinction law (A$_{\lambda}$ = A$_v$(0.55$\mu$m/$\lambda$)$^{1.6}$; Rieke \& Lebofsky 1985).  The results of this analysis are presented in the first two columns of Table 4.  The Q(2)/S(0) ratio is extremely sensitive to variations in line flux, uncertainties of $\sim$10\% in each feature can cause derived A$_v$ differences of more than 20 mag.  The uncertainty in the extinction is affected by imperfect telluric correction in the Q-branch emission region (2.40-2.44$\mu$m), and this was problematic for the spectra on T Tau, HL Tau and DG Tau.  HL Tau has an Q(3)/S(1) emission line ratio that is less than the intrinsic value, which suggests that the telluric correction was very poor and/or that scattered light could effect the extinction values (e.g., scattered light affects the spectra in the opposite sense from extinction, resulting in line ratios that are lower than the intrinsic value).  For further analysis of the line ratios, we dereddened the observed line fluxes using the extinction from the H$_2$ Q(3) and S(1) lines because these lines are intrinsically brighter than Q(2) and S(0) and are less sensitive to variation in the ratio (extinction toward HL Tau was assumed to be zero).

Figure 4 shows the population diagram of the ro-vibrationally excited H$_2$ derived from the detected emission features.  The level populations show that the H$_2$ is well described by thermal excitation with a single temperature in the 1800-2200K range (this is even roughly true for HL Tau, even though the Q-branch fluxes are affected by telluric correction).  This temperature range corresponds with the values derived by the early study of Burton et al. (1989), who observed a range of physical regions of known shock excitation.  Discussion of H$_2$ excitation mechanisms for the observed emission is continued in \S4.

The IFU spectra of T Tau has high S/N in several H$_2$ features which allows for the construction of emission line maps to study spatially resolved line ratios.  In addition to the continuum subtracted {\it v}=1-0 S(1) image (Figure 2), maps of the {\it v}=1-0 Q(3) and {\it v}=2-1 S(1) emission were also made for T Tau by integrating the over the H$_2$ features in continuum subtracted datacubes.  In Figures 5a-5c we show the integrated maps of the {\it v}=1-0 S(1) (5a), {\it v}=1-0 Q(3)(5b) and {\it v}=2-1 S(1) (5c) emission line fluxes in the vicinity of the T Tau triple.  For the {\it v}=1-0 S(1) and Q(3) images, the display scale is linear from -10\% to +90\% of the peak flux in the {\it v}=1-0 S(1) map.  To bring out low level emission, the {\it v}=2-1 S(1) image is scaled from  -10\% to +30\% of the peak flux detected in the {\it v}=1-0 S(1) map.  Overplotted in each of the emission line maps are the contours of the continuum emission from the stars in T Tau (as in Figures 1 and 2).  From these line flux maps we see that the global emission line structures are spatially similar for the different features.  The H$_2$ emission is brightest in the south toward the knots that encompass the position of T Tau S. 

In the map for the 1-0 S(1) emission line (Figure 5a) we identify three locations where we extracted spectra in 0.$''$2 diameter circular apertures.  These spectra were extracted to present examples of spatial locations in the IFU field that have emission lines with differing 1-0 Q(3)/1-0 S(1) and 2-1 S(1)/1-0 S(1) line ratios.  In Figures 6a-6c we present the three emission species extracted at the positions 1, 2 and 3 as defined in Figure 5a.  Each plot is scaled to a peak flux of 1.0 in the 1-0 S(1) line.  Position 1 has a  1-0 S(1):1-0 Q(3):2-1 S(1) ratio of 1.0:0.77:0.06.  Position 2 is 1.0: 0.81:0.19 and Position 3 is 1.0:1.1:0.05.  The Q(3) emission profile is $\sim$15\% broader than the 1-0 S(1) because of the longer wavelength of this feature.  Also, the Q(3) emission profile seems slightly blueshifted with respect to the 1-0 S(1) because this feature is close to the redward edge of the spectral range and the wavelength dispersion solution is less accurate than in the middle of the range.  However, the relative strengths of these emission lines clearly vary between the three positions where these spectra were extracted.

Following the analysis described previously in this section, a visual extinction was determined for each spectral pixel for the data on T Tau using the Q(3)/S(1) emission line ratios.  Figure 7a presents the map of the spatially resolved  Q(3)/S(1) emission line ratios.  The corresponding extinction structure determined from this analysis is included in grey in the figure key.  Contours of the Q(3) S/N are overplotted in black in Figure 7a.  To limit noise and the detrimental effects caused by telluric correction in the plot, only emission regions that had Q(3) flux with a S/N greater than 15.0 were included. Figure 7a shows that the derived line ratios correspond to visual extinction variations of more than $\sim$20 mag on spatial scales of less than 1$''$.  The median extinction derived over all spectral pixels is $\sim$9mag, which is higher than the typical values attributed to the system.  Yet, the 1$\sigma$ uncertainties in line ratios typically correspond to $\sim$5mag of A$_v$, so a median A$_v$ value of 9$\pm$5 is consistent within 2$\sigma$ to visual extinctions of $\sim$2mag toward T Tau N estimated in the literature.  Regions with the highest S/N in the Q(3) flux have average visual extinction of $\sim$4 mag, which is lower than the median value.  Variations of more than 10 magnitudes A$_v$ are seen in areas that have similar H$_2$ Q(3) S/N levels, which means that any variation in line ratio is not caused by poor or variable telluric correction of the Q(3) emission.  In the regions that exhibit line ratio variation with similar Q(3) signal, the S(1) line flux is changing by quite a bit.  This implies that the circumstellar material in the vicinity of T Tau is very clumpy.  In areas with higher estimated A$_v$ values, the dust is likely foreground to the observed H$_2$ emission.  We note that the absolute values for the extinctions derived from the H$_2$ line ratios are uncertain, since they assume that the obscuring material follows an ISM extinction law which may not be the case.  Deviations in the assumed power-law extinction to a more grey value can cause $\sim$10 mag differences in derived A$_v$s.  Nevertheless, the observed 1-0 Q(3)/1-0 S(1) line ratios clearly vary, which implies that strong spatial variations in obscuration likely exist toward this region.  The apparent enhanced extinction at the position of T Tau S (seen in white at x=$0.''2$ y=$0.''8$) is not real, investigation of individual spectra in this region show that the H$_2$ flux is weak and the Q(3) emission detection is affected by residual noise from the continuum subtraction process.

The map of spatially resolved 2-1 S(1)/1-0 S(1) line ratios for T Tau is presented in Figure 7b.  These values correspond to H$_2$ excitation temperature in the T Tau environment, and values for T$_{ex}$ are included in grey in the figure key.  The values in this figure were determined by dereddening the {\it v}=2-1 S(1)/{\it v}=1-0 S(1) line ratios at the position of each spectral pixel using the extinction presented in Figure 7a, and assuming LTE level populations to relate the line ratios to an excitation temperature.  Overplotted in black in Figure 7b are contours of the 2-1 S(1) emission.  Peaks in the S/N of the 2-1 S(1) emission are consistent with the Q(3) and S(1), though deviations are apparent such as the higher relative strength of the knot of 2-1 S(1) emission at x=1.$''$5 y=0.$''$25 in Figure 7b. The color scale in Figure 7b represents the line ratios and corresponding excitation temperatures.  The median value is 2320$\pm$100K, and variations in T$_{ex}$ of less than 300K are seen in over 70\% of the field (green in Figure 7b).  Areas with low variation in T$_{ex}$ include regions with both high and low H$_2$ fluxes and extinction changes of 5-15 magnitudes.

There are regions in the map of excitation temperature for T Tau (Figure 7b) that have 0.$''$1-0.$''$2 spatial extents where the T$_{ex}$ varies by more than $\sim$800K from the median level.  The previously mentioned knot of high relative 2-1 S(1) emission to the south of T Tau S has an average T$_{ex}$ of $\sim$3100K.    Several regions of low excitation, $\sim$1400K, are also seen in the area $\sim$0.$''$6 to the south and west of the position of T Tau S.  The highest excitation temperatures of $\sim$4000K are found directly to the north and south of the position of T Tau S within $\sim$0.$''$3 ($\sim$42 AU).  Higher excitation temperatures determined from the 2-1 S(1)/1-0 S(1) line ratios are sometimes associated with UV fluorescent excitation and deviations from LTE.  The regions with T$_{ex}\sim4000$K may provide evidence for emission arising from UV stimulation within $\sim$50 AU of T Tau South. We note that our analysis of the H$_2$ excitation temperatures necessarily assume that the H$_2$ detected at a given position in the IFU field is isothermal, so the line ratios along each spatial line of sight trace gas with a single temperature rather than a range of temperatures.  Comparable analysis of spatial structure in the 1-0 Q(3)/1-0 S(1) and 2-1 S(1)/1-0 S(1) line ratios from the observations of HL Tau and HV Tau C will be included in forthcoming publications.

\subsection{Kinematics of the H$_2$ Emission}

To test for kinematic structure seen in the H$_2$ emission, the velocity barycenter, v$_c$, and dispersion, $\sigma_{v}$, were calculated.  These values are useful for measuring kinematic structure in emission features and are defined as:  

\begin{equation}
v_c = \frac {\int v I_v dv}{\int I_v dv}
\end{equation}

\begin{equation}
\sigma_v^2 = \frac {\int (v-v_c)^2 I_v dv}{\int I_v dv}
\end{equation}
where I$_v$ is the emission intensity at velocity, v (Riera et al. 2003).  These values are effectively the velocity moments of the emission line profiles.  The barycenter velocities and dispersions for a single sky emission line were calculated to test the accuracy of using these values to trace kinematic structure in NIFS IFU data.  The sky emission line used for this analysis was centered at 2.0725$\mu$m and had a S/N above the background of $\sim$12.  The sky line strength was consistent with the minimum S/N adopted for detection of H$_2$ emission in the YSOs.  Figures 8a and 8b present the spatially resolved v$_c$ and $\sigma_{v}$ values for the sky line and provide a measure of the accuracy and stability of these calculations for the IFU data.  The mean v$_c$ is 0.3$\pm$2.6km/s, and the $\sigma_{v}$ is 30.2$\pm$3.2km/s, where the uncertainty values are derived from the multiple measurements over all spectral pixels.  These values for v$_c$ and $\sigma_v$ are included in the top line of Table 5.  The figures are plotted with color scales that range from $\pm$3 standard deviations from the median value of v$_c$, and from 0 to twice the median value of $\sigma_v$.

The v$_c$ and $\sigma_{v}$ values for the {\it v}=1-0 S(1) emission were determined for each of the six CTTSs at every spatial pixel that had an H$_2$ emission greater than S/N=12.0.  The limit in S/N was chosen to decrease effects from noise in regions of low emission flux.  Figure 9 presents the maps of v$_c$ for each star, and the third column of Table 5 lists the median H$_2$ emission velocities and their standard deviations found from the multiple measurements over all spectral pixels.  The barycenter velocities were corrected for heliocentric effects and are referenced to the systematic radial velocity (included in the second column of Table 5; Hartmann et al. 1986; White \& Hillenbrand  2004).  The color scales in Figure 9 represent velocities at $\pm$3 standard deviations from the median for each star.  

The median values for v$_c$ deviate from the systematic radial velocity by a significant amount for HV Tau C, T Tau and RW Aur, while DG Tau, HL Tau, and XZ Tau have H$_2$ v$_c$ values within less than 10km/s of the radial velocity of the parent stars.  The H$_2$ emission is seen to be blue-shifted (or slightly blue-shifted) for all stars except for RW Aur.  The collimated, red-shifted jet from RW Aur is obvious from the H$_2$ velocity map, barycenter velocities of over 100km/s are seen to the left of the star in Figure 9.  T Tau shows lower relative v$_c$ values in the south with respect to the north, which is largely consistent with the known blue and red-shifted lobes of the wide angle outflow associated with HH 255 (Solf \& B\"ohm 1999).

In all cases, the standard deviations of multiple measurements of v$_c$ are greater for H$_2$ emission in the CTTSs than for the sky line, which suggests that velocity shifts are present beyond those explained by systematic effects in the IFU.  Though, the difference in standard deviation is at a low level of significance for HL Tau and XZ Tau.  For several of the stars (i.e., DG Tau, HV Tau C, HL Tau), the velocity structures seen in Figure 9 show that regions of high H$_2$ S/N have lower variations in v$_c$ and appear mostly in green at the median level.  The maps of v$_c$ for DG Tau and  HV Tau C have moderate velocity shifts (15-20 km/s) in consecutive pixels in regions with lower S/N.  Investigation of the data suggests that these shifts are real in the data.  The shifts could arise from effects of lower S/N in the spectra or from spatially resolved, clumpy H$_2$ emission in the vicinity of the young stars which unevenly illuminate the 0.''1 width of the NIFS spectral slit and cause corresponding velocity barycenter shifts (Bacciotti et al. 2002).

The velocity dispersion values, $\sigma_v$, are listed in the fourth column of Table 5 and are presented in Figure 10.  The color scale in Figure 10 is plotted from 0.0 to two times the standard deviation for each star.  For all except RW Aur, the measured velocity dispersions are consistent with the value determined from the sky line, and hence with spectrally unresolved H$_2$ emission.  RW Aur has a median velocity dispersion that is nearly double the sky value, and it is the only star in the sample which exhibits broad H$_2$ emission line profiles arising from high and low velocity gas.  HV Tau C, T Tau and DG Tau all have standard deviations in the dispersion that are greater than for the sky line.  Some areas in the maps for HV Tau C and DG Tau show velocity dispersions that are less than the instrumental value, but most of these correspond to areas that also have pixel-to-pixel structure in v$_c$.  Pixel-to-pixel deviations in $\sigma_v$ and dispersions less than the instrumental value are consistent with the interpretation that the H$_2$ arises from spatially resolved regions of slightly clumpy, discontinuous emission that unevenly illuminate the NIFS slit width.

T Tau and RW Aur are the only two stars which exhibit statistically significant structure in their H$_2$ barycentric velocity profiles (Figure 9).  Further investigation of the H$_2$ data show clear velocity shifts in the spatial location of H$_2$ emitting regions.  Figures 11 and 12 present the H$_2$ {\it v}=1-0 S(1) velocity channel maps for T Tau and RW Aur, respectively.  For T Tau (Figure 11), the three channels are single wavelength slices through the H$_2$ datacube, and for RW Aur (Figure 12) the three panels are averages of two velocity slices per channel.  Figure 11 shows that many of the southern knots of H$_2$ emission that encompass the position of T Tau S are bright at all wavelengths through the velocity extent.  However, there is an increase in blue-shifted emission to the south of T Tau, and emission in the northern regions of the IFU field is seen only in the 0.0km/s and +29.7km/s channels.  On average, the velocity structure of H$_2$ emission is found to vary from slightly blue-shifted in the south to slightly red-shifted by $\sim$20km/s in the north of the IFU field.  In Figure 12, H$_2$ emission in the inner environment of RW Aur is seen in the low velocity channel (Figure 12a).  This low velocity gas encompasses the star and exhibits a linear extension to the east at a $\sim$120$^{\circ}$ position angle from the red-shifted jet.  Figure 12 shows that the H$_2$ emission from RW Aur becomes more red-shifted with increasing distance from the star toward the northwest.  The very high velocity emission from RW Aur extends to greater than 120km/s and more than 140 AU (1$''$), but little of this flux arises from the position of the star.

\section{Interpretations}

Molecular hydrogen emission from the vicinity of T Tauri stars is believed to arise from both quiescent gas in circumstellar disks (stimulated by UV/X-ray heating or from UV pumping) as well as from shock excited molecular gas in YSO outflows.  All six of the stars observed for this study are known to drive HH energy flows, but several are also known to have cool, dense gas indicative of substantial reservoirs of circumstellar disk material.  Our goal with this study is to identify the near IR H$_2$ emission mechanisms to better understand the origin of the gas, and relate the detected H$_2$ to disk or outflow characteristics detected at other wavelengths. 

Several of the T Tauri stars studied here have H$_2$ emission seen in UV electronic transitions pumped by Ly$\alpha$ flux (Ardila et al. 2002; Saucedo et al. 2003; Walter et al. 2003; Herczeg et al. 2006), so it seems plausible that some of the near IR H$_2$ we detect is from the corresponding ro-vibrational cascade of emission initially excited by this UV fluorescence.  The small range of T$_{ex}$ values of 1800-2200K determined from the H$_2$ level populations (Figure 4) suggests that most of the emission we see arises from a thermalized gas.  The observed line ratios of 2-1 S(1)/1-0 S(1) emission in all six of the stars are also characteristic to shock excited regions (Burton et al. 1989; 1990 Eisloffel 2000; Davis et al. 1996; 1999).  However, this line ratio diagnostic cannot exclude UV fluorescence conclusively because regions of high gas density or high UV flux can affect the population of high vibrational levels in the IR and will result in thermalized populations (Sternberg \& Dalgarno 1989; Burton et al. 1990; Smith et al. 1995).  Additionally, within the inner $\sim$30AU of the central stars Ly$\alpha$ emission is often very broad and is efficient at pumping the UV Lyman-Werner bands of UV H$_2$ (Ardila et al. 2002; Walter et al. 2003; Herczeg et al. 2004).  Hence, for all stars in our study, it is not possible for us to distinguish conclusively between UV Ly$\alpha$ pumping of high density gas and shock excitation as the main H$_2$ stimulation mechanisms in the inner regions of the stars.  Takami et al. (2004) ruled out UV fluorescence as a viable H$_2$ excitation mechanism for DG Tau based on the large mass momentum fluxes that resulted from their calculations.

Table 6 presents molecular hydrogen line ratios from four sets of transitions detected in our spectra which were chosen because they vary depending on different H$_2$ excitation mechanisms.  The second column shows the expected line ratios from calculations of low density gas with UV fluorescent emission by Black \& van Dishoeck (1987), and the following six columns present the measured values from our spectra.  The line ratios for the UV fluorescent emission are dissimilar to the values we measured in the CTTSs, and they show particular deviation for the {\it v}=2-1 S(1)/{\it v}=1-0 S(1) ratio.  The {\it v}=2-1 vibrational levels are not populated to the extent predicted by UV fluorescence of quiescent low density gas.  If UV stimulation from the central star is significant in the inner $\sim$50AU environs of these CTTSs, the H$_2$ emission must arise from high density, thermalized gas.

Nomura \& Millar (2005) and Nomura et al. (2007) have predicted that molecular hydrogen heated by stellar UV and X-ray emission should cause maximum gas temperatures less than $\sim$1000K at distances greater than $\sim$30-50 AU from the star.  Thus, stellar UV or X-ray heating should produce neither the observed H$_2$ level populations with LTE excitations of $\sim$2000K nor the strong H$_2$ flux we find at greater than $\sim$30AU distances from the stars.  These models also imply that the H$_2$ should be centered on the parent star and decrease with increasing distance.   Of the six stars in our sample, only DG Tau and HV Tau C have H$_2$ emission that decreases with distance from the star.  In both of these cases, the H$_2$ has a morphology that seems to trace an inclined circumstellar disk (Figure 2), and it extends to greater than 100AU distances.  UV and X-ray heating of the quiescent disk material is a viable excitation mechanism for emission in these two stars only if the current models underestimate the excitation temperatures and spatial extents of the H$_2$ by factors of 2-3 or more (Nomura \& Millar 2005; Nomura et al. 2007).  This seems unlikely since quiescent heating of material by the central star cannot easily explain H$_2$ gas at excitation temperatures of greater than 1800K beyond 50AU distances.   Recent observations have also shown that X-ray emission can arise from shocks within the outflows themselves (Favata et al. 2006; G\"udel et al. 2005), so spatially extended H$_2$ could conceivably be heated locally by X-ray induced shocked emission.  Of the stars in our sample, only DG Tau has known X-ray emission in its outflow to date (G\"udel et al. 2005).
 
For both DG Tau and HV Tau C, the detected {\it v}=1-0 S(1) H$_2$ emission decreases with increasing distance from the central continuum flux.  Figure 13a and b present the spatial profiles of the H$_2$ emission from DG Tau and HV Tau C plotted versus the distance from the position of the central star (approximated for HV Tau C as the center of the dark lane in the scattered light nebulosity).  The H$_2$ is only spatially averaged over the bright flux from the blue-shifted (lower right in Figure 2) and red-shifted (right in Figure 2) lobes of the emission for DG Tau and HV Tau C, respectively.  Overplotted in Figures 13a and b are the average shapes of the continuum emission and, for comparison, a curve that follows a 1/r$^2$ power law decrease in flux with increasing distance (scaled to the H$_2$ emission level at 0.$''$24 distances from the stars).  The H$_2$ flux decrease with increasing distance from DG Tau and HV Tau C resembles (but does not exactly correspond to) a 1/r$^2$ decrease.  The observed H$_2$ emission from HV Tau C does not follow the structure seen in atomic emission in the known HH outflow, which shows bright knots at extended distances from the star (Stapelfeldt et al. 2003; Beck et al. in prep).  However, the H$_2$ from DG Tau does resemble the low velocity, smoothly decreasing, wide-angle atomic emission seen in the HH 158 outflow (Bacciotti et al. 2002; Takami et al. 2004).  The most likely explanation for the detected H$_2$ in the spatially extended ($>$50 AU) regions around DG Tau and HV Tau C is from shock excited emission from a wide, slow component of the outflow (wind) that encompasses the collimated jets from the stars (Takami et al. 2004).

  The spectra for T Tau, HV Tau C and RW Aur show statistically significant shifts from the systemic radial velocity in H$_2$, which is a clear indicator of emission arising from shocks in the outflows (Figures 11, 12).  Discrete knots of H$_2$ at spatially extended distances from the central stars are also suggestive of an origin in shock excited emission (e.g. HL Tau, XZ Tau).  Three types of shock excited H$_2$ were defined and described in detail by Kwan (1977), Draine (1980) and Draine, Roberge \& Dalgarno (1983).  Non-magnetic shocks (pure jump shocks) cannot explain T$_{ex}$ of greater than $\sim$500K because H$_2$ is dissociated in the shock front and the gas cools by pure rotational emission in the post-shock regions (Kwan 1977).  Dissociative shocks with magnetic precursors, or jump shocks (J type), are named as such because the magnetohydrodynamical parameters are discontinuous across the shock front.  Continuous shocks (C type) have continuous parameters across the shock, they arise in regions with stronger magnetic fields, and the magnetic cushioning results in little (or no) H$_2$ dissociation across the shock (Draine 1980; Drain, Roberge \& Dalgarno 1983).

 The last two columns of Table 6 present emission line ratios for J and C type shock calculations from Smith (1995).  Comparison of these model ratios with our data in the preceding columns of Table 6 reveal that the C type shock calculations are consistent with the H$_2$ line ratios in the observed spectra.  Aside from a factor of two difference in the observed 2-1 S(1)/1-0 S(1) flux, the line ratios for RW Aur correspond almost precisely to those presented for the C shock model.  HL Tau, which is likely observed strongly in scattered light or affected by poor telluric correction, shows deviations from both shock models.  Variations in the model assumptions for the J and C type shocks affect the resulting calculations, and we note that the C35 and J8 models from Smith (1995) are both consistent with the 2-1 S(1)/1-0 S(1) line ratios detected in our spectra.  Our goal here is not to argue details of J and C type shocks applied to these systems, but to strengthen the argument that the H$_2$ emission we detected at spatially extended distances from the CTTSs is excited in shocks in the outflows rather than primarily from UV or X-ray excitation by the central star. 

To date, most of the studies that analyze differences in shock models are based on observations of outflows propagating into the ambient ISM at distances of many arc seconds from the parent star (100s to 1000s of AU).   Moreover, the level of H$_2$ arising from shocks depends very strongly on parameters such as the local magnetic field strength and the local ionization fraction in the ambient material.  Typical magnetic field strengths considered by the shock calculations are $\sim$1 milliGauss or less.  Though, measurements of the magnetic fields from Zeeman split absorption features in CTTSs have shown that strong fields of more than 1 kG exist near to the star (Johns-Krull et al. 1999; Valenti \& Johns-Krull 2004; Yang et al. 2005; Daou et al. 2006).  Ray et al. (1997) also found spatially resolved gyrosynchrotron emission in the vicinity of T Tau S, suggesting magnetic fields of several Gauss exist in the outflows at spatially extended distances from the stars.  Bacciotti \& Eisloffel (1999) also showed that the ionization fraction decreases with increasing distance from a star.  Although the C type shocks seem to agree with line ratios for the H$_2$ emission in our observed stars (Table 6), the shock model values for the observed H$_2$ line ratios are calculated for outflows propagating into ISM material.  Hence, the models for H$_2$ shocks may not use accurate assumptions or physical parameters for the inner $\sim$100 AU of T Tauri stars where magnetic fields and ionization fractions are likely greater than ISM levels (Hartigan et al. 2007; Bacciotti \& Eisloffel 1999).

\section{H$_2$ from the Individual Stars}

\subsection{DG Tau}   The molecular hydrogen {\it v}=1-0 S(1) in the vicinity of DG Tau, presented in Figure 2, shows a v-shaped morphology (lower right), a ``dark lane'', and a ridge of emission toward the left side of the star.    The bright, v-shaped morphology encompasses the inner blue-shifted HH 158 outflow (Dougados et al. 2000; Takami et al. 2004).   The observed structure in the H$_2$ emission is reminiscent of morphologies of circumstellar disks seen in scattered light toward embedded proto-stars (e.g. Padgett et al. 1999).  We interpret the dark lane in the H$_2$ emission as optically thick material in the spatially extended accretion disk around DG Tau observed inclined to the line of sight (Testi et al. 2002; Pyo et al. 2003).  This dark lane of no H$_2$ flux is aligned with the red and blue-shifted lobes of millimeter emission from the rotating circumstellar disk (Testi et al. 2002).  

Along the outflow axis (oriented at 226$^{\circ}$), the H$_2$ emission extends to a projected distance of $\sim$0.$''$9 ($\sim$126 AU assuming a 140 pc distance to Taurus) on each side of the star.  Perpendicular to the outflow axis, the H$_2$ emission is nearly symmetric with projected extents up to 0.$''$5 ($\sim$70 AU).  These extents in the H$_2$ emission are slightly greater than those measured by  Takami et al. (2004), but this is likely a result of longer exposure time and higher S/N in our data.  The brighter lobe of H$_2$ emission also shows similarities to the low velocity component of [S II] atomic emission measured by Bacciotti et al. (2002).  Takami et al. (2004) first proposed that the shock excited H$_2$ arises from a wide-angle, warm molecular wind that surrounds the inner regions of the HH 158 outflow. H$_2$ emission from UV electronic transitions studied by Ardila et al. (2002) and Herczeg et al. (2006) were consistent with this interpretation, and the data presented here is also in agreement.  We find that a molecular wind of shocked H$_2$ emission from DG Tau decreases with increasing distance from the star (Figure 13) and extends along the outflow axis to distances of $\sim$130 AU.  

\subsection{HL Tau}  The collimated jet from HL Tau was discovered by Mundt, Ray \& Buhrke (1988) in their survey for optical jets from the six YSOs in the HL Tau region.  Molecular hydrogen from HL Tau is detected to the northeast and southwest of the star (Takami et al. 2007), but the H$_2$ is stonger to the northeast in the region of the blue-shifted outflow.  The H$_2$ emission to the southwest is detected at a low relative S/N and is not studied in this work (Takami et al. 2007).  In Figure 2, the axis of the HH 150 outflow from HL Tau was aligned parallel to the x direction of the field.  The centers of the two bright clumpy structures of H$_2$ emission are offset from the axis of the collimated atomic emission (Pyo et al. 2006; Takami et al. 2007).  This result may be consistent with previous observations and shock models which show that slow, magnetic shocks with H$_2$ cooling encompass the faster on-axis atomic emission from dissociative shocks (Davis et al. 2000; 1999; 1996).  However, from our data the line ratios seem to be adversely affected by levels of scattered light and do not clearly correspond to the shock ratio calculations (Smith 1995).   

In the inner environment of HL Tau, Pyo et al. (2006) showed that the [Fe II] emission decreases significantly toward the star, with little or no flux arising within $\sim$ 0.$''$8 of the central star.  We also find a decrease in H$_2$ emission strength toward the position of the star within 0.$''$4, and the observed H$_2$ morphology is consistent with enhanced obscuration from material in the circumstellar disk proposed by Pyo et al. (2006) and viewed previously in scattered light and millimeter emission (Lucas et al. 2004; Close et al. 1997; Mundy et al. 1996; Wilner et al. 1996).  Further discussion of these data are included in a companion publication (Takami et al. 2007)

\subsection{HV Tau C}  Like the H$_2$ from T Tau, the molecular hydrogen emission from HV Tau C extends over 60\% of the IFU field at a S/N greater than 12.  However, the emission from HV Tau C does not exhibit bright, discrete knots that are spatially extended from the continuum (Figure 2).  The molecular hydrogen in the vicinity of HV Tau C is instead continuous and decreases in intensity with increasing distance from the bright continuum flux (Figure 13b).  Additionally, the emission also shows the dark lane seen in scattered light images and traces the circumstellar disk viewed edge-on (Monin \& Bouvier 2000; Stapelfeldt et al. 2003).

The HV Tau C observations were acquired so that the outflow axis was parallel to the x direction in Figure 2.  The blue-shifted region of the outflow extends to the left in the figure center, and the red-shifted jet is to the right (Beck et al. in prep).  The molecular hydrogen shows no flux increase or decrease along the outflow axis and no obvious correlation with the atomic emission from the HH 233 outflow (Stapelfeldt et al. 2003).  The H$_2$ level popultions are consistent with warm, thermally excited gas with a temperature of 2030K.  Like the H$_2$ from DG Tau, the most plausible explanation for molecular hydrogen detected in HV Tau C is from a wide angle warm molecular wind that encompasses the HH 233 outflow.  Under this assumption, the wind extends to greater than 50 AU distances from the approximate position of the parent star (in the center of the field; Figure 1).  Further discussion of these data and presentation of the warm molecular wind in reference to the atomic emission in the outflow will be included in an upcoming publication (Beck et al., in prep.).

\subsection{RW Aur}  The H$_2$ from the vicinity of RW Aur is brightest within 70 AU (0.$''$5) of RW Aur A, but it shows a linear extension to the northwest (left in Figure 2, Figure 12) and another, weaker extension to the east.  The observation of RW Aur was made with the outflow axis parallel to the x direction in the NIFS field.   The extension of emission to the east is at $\sim$+10km/s with respect to the radial velocity of the system (Figure 9; 12).  This emission is likely associated with the low velocity component (LVC) studied  by Pyo et al. (2006).  It is difficult to interpret where this low velocity extension of H$_2$ emission to the east arises from, since it is misaligned with the red and blue-shifted outflows.  Hence, the emission does not exhibit a morphology that encompasses the flow which could imply an origin in a wide-angle wind (e.g. DG Tau, HV Tau C).  The total exposure time on RW Aur was comparably short (400sec; Table 1), and higher S/N data would be useful to investigate this emission component.

The morphology of the H$_2$ emission that extends to the northwest of RW Aur is consistent with the inner regions of the known HH 229 outflow.  We detect only the red-shifted jet in H$_2$, which is consistent with the curious result of past studies that show that the red-shifted atomic emission is brightest (Woitas et al. 2005; Pyo et al. 2006).  The jet emission is red-shifted to $>$100km/s from the systemic radial velocity at distances of $\sim$140 AU (1$''$; Figure 7; Figure 12).  RW Aur is the only star for which we see strongly shifted shock velocities in the H$_2$ emission, and the kinematics of the H$_2$ are similar to the [S II] and [Fe II] atomic species studied by Woitas et al. (2005) and Pyo et al. (2006).  Additionally, RW Aur has a higher median velocity dispersion (Figure 10) because the H$_2$ line profiles in the inner regions arise from two components, the HVC and LVC (Pyo et al. 2006). The LVC component is stronger in total flux within $<$70 AU of the star, but it also shows the extension to the east.  So, the HVC and LVC emissions are not spatially coincident in all regions (Davis et al. 2001; 2002).  The line ratios of multiple H$_2$ features in the region of brightest emission (mainly arising from the LVC) seem consistent with an origin in C type shocks (Table 6; Smith 1995).

\subsection{T Tau}  The H$_2$ emission detected in T Tau shows complicated structure, with the brightest regions of emission to the south and west that encompasses T Tau S in a loop-like morphology.  A ``knots+arcs'' structure in H$_2$ emission was detected to $\sim15''$ distances from T Tau by Herbst et al. (1996; 1997).  Additionally, loops of atomic emission were found by Robberto et al. (1995) to the south of T Tau S in the region of HH 255 (Burnham's nebula).  Solf \& B\"ohm (1999) showed that the atomic emission species in the T Tau region trace the HH 155 east-west outflow from T Tau N, and the HH 255 north-south outflow from T Tau S.  The lobes of emission in [S II] show a redshift to the east and north of the system from the two respective outflows, and blue-shifted emission regions lie to the west and south (Solf \& B\"ohm 1999).  Comparison of our H$_2$ map with the atomic data suggests that the bright emission to the south and west of T Tau S could correspond to a blue-shifted lobe of the HH 255 outflow.  The velocity maps presented in Figures 9 and 11 show an average $\sim$20km/s shift between the regions of emission to the north of T Tau N and the brighter areas to the south and west of T Tau S.  While the character of the velocity shifts is similar, the kinematics of the H$_2$ are on a lower velocity scale than the [S II] (Solf \& B\"ohm 1999).  This could be consistent with observations that show that molecular hydrogen arises from slower wings of the molecular bow shocks rather than the on-axis flow traced by atomic emission (Davis et al. 1999; 2000; Eisloffel et al. 2000).  The outflow morphology seems consistent on a rough scale to that shown by Solf \& B\"ohm (1999).  However, detailed comparison with their figures is difficult because the spatial resolution of our data is nearly an order of magnitude higher.

The relation between ro-vibrational H$_2$ in the inner regions of T Tau and emission from electronic transitions in the UV is investigated by comparing our map of H$_2$ emission to data acquired by Saucedo et al. (2003) and Walter et al. (2003).  These UV studies found that emission in the inner vicinity of T Tau is consistent with H$_2$ fluorescently pumped by Ly$\alpha$ photons (likely both from the central stars and from shocks in the outflows).  UV stimulated fluorescent H$_2$ can show excitation temperatures of upwards of 4000K determined from the 2-1 S(1)/1-0 S(1) line ratio (Black \& vanDishoeck 1987).  Figure 7b reveals regions near the position of T Tau S that have T$_{ex}$ of $\ge$4000K based on the 2-1/1-0 S(1) line ratio, and this H$_2$ could also conceivably arise from low density gas where the vibrational temperature is determined by UV pumping (Black \& van Dishoeck 1987).  Curiously, by comparison Walter et al. (2003) found a pronounced decrease in the UV H$_2$ flux at the nominal position of T Tau S.  

Away from the position of T Tau S, the thermally populated line ratios and character of the H$_2$ that we find in the infrared (Figure 7b) are not consistent with excitation by the central stars, but could still be UV pumped by the significant Ly$\alpha$ photons aring from shocks in the outflows (Walter et al 2003).  In their Figure 4, Saucedo et al. (2003) presented a map of the areas with strong UV H$_2$ emission.   The bright arc of {\it v}=1-0 S(1) emission that we see to the southwest of T Tau N and west of T Tau S corresponds to the location of the ``N1'' emission identified by Saucedo et al. (2003).  Additionally, the ``S1'' region of emission that they identify to the south-southeast of T Tau S also corresponds to the lower area of the bright IR H$_2$ we see at roughly this same location.  Our respective observations were carried out 5 years apart, so unambiguous identification of corresponding H$_2$ knots in these different wavelength regions is difficult because of possible time variability of features.  Also, the levels of extinction along the line of sight will affect the UV H$_2$ emissions more strongly than the near IR lines.

Comparison of multi-epoch observations obtained with similar spatial coverage at corresponding wavelengths allow us to investigate the stability of H$_2$ emission structures in the vicinity of T Tau.  Thus, we compare our map of {\it v}=1-0 S(1) emission in Figure 2 to the adaptive optics fed high spatial resolution map of H$_2$ in Figure 5 presented by Herbst et al. (2007).  Herbst et al. also detect the ``N1'' knot of emission to the southwest of the position of T Tau N that we see, so this flux seems to have a stable location throughout the duration of high resolution spatially resolved H$_2$ observations (Saucedo et al. 2003).  Overall, the ``loop'' of H$_2$ emission found surrounding the position of T Tau S by Herbst et al. (2007) is similar to the H$_2$ we detect, but there are obvious differences.  The ``S1'' or ``S2'' knots identified by Saucedo et al. (2003) are not clearly found in their images.  Additionally, in their Figure 5, Herbst et al. find a knot of H$_2$ $\sim$0.$''$2-0.$''$3 to the south of the T Tau S binary that is fainter than the two knots of H$_2$ that they find to the west of these stars.  We also detect a knot of emission to the southeast of T Tau S, but it is 0.$''$4-0.$''$5 away from the position of the stars and it includes the brightest clump of spatially extended H$_2$ emission that we detect (e.g. see our Figure 5).   The brightest knot of {\it v}=1-0 S(1) emission that is seen by Herbst et al. (2007) within $\sim1''$ of the stars is their ``C2'' knot, which corresponds to the location of the forementioned ``N1'' knot of emission near T Tau N.  The H$_2$ imaging spectroscopy of Herbst et al. (2007) was acquired in Dec. 2002.  Hence, by comparing their data with our H$_2$ images we find that the spatial locations and brightnesses of the H$_2$ emission knots in the inner $\sim$1$''$ of the position of T Tau have varied on $\sim$3 year timescales between our respective observations.

The N1 emission in the environment of T Tau N seems to have a stable location during the $\sim$5 year timescale over the course of high spatial resolution H$_2$ observations of the system (Saucedo et al. 2003; Herbst et al. 2007; Figure 5).  In the inner environs of T Tau, it seems plausible that stationary shocks are defined by the N1 position where material in the outflow plows into the walls of the existing outflow cavity (Saucedo et al. 2003; Stapelfeldt et al. 1998).  However, if the N1 knots of H$_2$ emission trace stationary shocks in a cavity wall then the geometry of circumstellar material in the T Tau system may be more complicated than is presently known.  In YSO multiples, disks (and presumably outflow cavities) should be truncated by tidal effects at distances of less than $\sim1/3$ of the orbital semi-major axis of the system (Artymowicz et al. 1991).  However, spatial stability of the N1 knots of H$_2$ implies that significant material exists near T Tau at a projected distance comparable to the projected separation of the T Tau N / T Tau S system.  

\subsection{XZ Tau}  XZ Tau lies just $\sim$20$''$ away from HL Tau, in a mini-cluster of six YSOs in a 0.05$\times$0.1pc spatial region.  The initial discovery of outflows of material from XZ Tau was made in the optical study of Mundt, Ray \& Buhrke (1988).  The H$_2$ emission we find toward XZ Tau shows an arc-like morphology extending from the southeast to the southwest of the 0.$''3$ separation XZ Tau AB binary system.  The structure in the H$_2$ looks like an incomplete ``ring'' of emission, and resembles the scattered light disk seen to the south of the GG Tau A binary (Roddier et al. 1996).  Of the stars observed in this sample, the kinematics of H$_2$ in the vicinity of XZ Tau are the most consistent with expected emission from a quiescent disk - the H$_2$ is centered on the systemic radial velocity with little deviation in velocity centroid and dispersion compared to instrumental values measured by the sky line.  The molecular hydrogen extends to more than a projected distance of 80 AU from the stars, and the level populations indicate warm gas at a temperature of $\sim$2270K.

  The high resolution HST images of Krist et al. (1997) and Coffey et al. (2004) showed expanding bow shocks of emission extending to greater than 5$''$ distances north of the XZ Tau binary, but a lower level of atomic emission was detected to the south.  These studies also revealed that the outflow from XZ Tau is poorly collimated and exhibits bow shocks with large, $\sim3''$ spatial extents within less than 10$''$ distances from the stars.   Based on the energetics and the known extent of outflows from the system, the arc of emission to the south of XZ Tau AB most likely arises from a bow shock associated with the inner, southern HH 152 outflow.  Future observations to monitor the tangential motions of the H$_2$ with respect to the known atomic outflow emission could clarify the nature of molecular hydrogen in this system.

\section{Summary}

We have presented adaptive optics integral field spectroscopy of molecular hydrogen in the inner 200 AU environments of Classical T Tauri Stars using the Near-infrared Integral Field Spectrograph at Gemini North Observatory.  The key findings from our study are:

1) Spatially extended {\it v}=1-0 S(1) (2.12$\mu$m) molecular hydrogen emission is detected in six Classical T Tauri Stars that drive Herbig-Haro energy flows.  For HL Tau, HV Tau C, RW Aur and XZ Tau these observations present the first spatially resolved detections of near IR H$_2$ in the vicinity of the young stars (Takami et al. 2004; Duchene et al. 2005).   The majority of the H$_2$ emission detected in our study is not coincident with the continuum flux from the stars.

2) In several cases, the morphologies of the H$_2$ emission are consistent with the emisison from the known HH energy flows.  Particularly, RW Aur, HL Tau and T Tau roughly show H$_2$ that follows the location and knots+arcs morphologies seen in atomic emission species from the outfows.
 
3) In the K-band region, extraction of spectra using small apertures from integral field spectroscopic data increases the detection sensitivity to weak H$_2$ emission lines over the continuum flux.  In addition to the {\it v}=1-0 S(1) (2.12$\mu$m) feature, we detect up to 8 other H$_2$ transitions in the spectra of the stars. 

4)  Line ratios of {\it v}=1-0 Q(3)/{\it v}=1-0 S(1) and {\it v}=1-0 Q(2)/{\it v}=1-0 S(0) are used to estimate extinctions toward each of the stars, assuming attenuation by material which follows an ISM extinction law.  2-D maps of line ratios are used to spatially resolve the extinction and excitation temperature in the vicinity of the T Tau triple star.  We find large variations in line ratios, and hence extinction and excitation temperature, on spatial scales of less than 100 AU in the environment of T Tau.

5) The level population diagrams derived from the multiple H$_2$ features show that the emission is consistent with gas in thermal equilibrium with temperatures in the 1800K to 2300K range.

6)  The centroid velocities and velocity dispersions of the {\it v}=1-0 S(1) emission line are determined for all stars in spectral pixels that had S/N greater than 12.0  (Figures 9 and 10). The resulting maps of the spatially resolved velocity and dispersion structure were compared to the calculations carried out on a single sky emission line (used to estimate the instrumental performance).  Three of the six stars show shifts with respect to their systemic radial velocities, and an additional two show velocity structure in the velocity dispersion or standard deviations.  RW Aur has a H$_2$ profile that is very broad because of the contribution of high velocity emission, and the H$_2$ kinematic structure is consistent with the known atomic emission from its HH energy flow.  XZ Tau is the only star which shows velocity and dispersion kinematics consistent with spectrally unresolved H$_2$ emission centered at the stellar radial velocity.

7)  For all six stars, the observed spatial extents and excitation temperatures of the H$_2$ are factors of $\sim$2-3 or more than predicted by models of UV or X-ray heating by the parent star of quiescent gas in circumstellar disks.  Additionally, the H$_2$ spatial extent and the ro-vibrational level populations are not consistent with emission exclusively from FUV or Ly$\alpha$ stimulated fluorescence arising from the central star.  

8) The observed H$_2$ emission line ratios are compared to theoretical values calculated for UV excited fluorescent emission (Black \& van Dishoeck 1987) and for J and C type shock models (Smith 1995).  The H$_2$ line ratios are found to be most consistent with shock excitation for all cases but HL Tau, which appears to have line fluxes that are adversely affected by strong levels of scattered light.  However, these models for UV fluorescent emission and shock excited H$_2$ emission were generated for ISM environments that have different physical conditions than typically found within 200 AU of Classical T Tauri Stars.

9) Taken in combination, the H$_2$ emission morphologies, detection of molecular hydrogen at distances of greater than $\sim$50 AU from the star, excitation temperatures of more than 1800K, kinematics of the features, and the consistency with existing shock models combine to indicate that the bulk of the H$_2$ in these stars arises from shock excited emission associated with the known HH energy outflows rather than from quiescent H$_2$ disk gas stimulated into emission by flux from the central star.



\acknowledgments

We are extremely grateful for the support of the NIFS teams at the Australian National University, Auspace, and Gemini Observatory for their tireless efforts during the instrument integration, commissioning and system verification: Jan Van Harmleen, Peter Young, Mark Jarnyk (deceased), Nick Porecki, Richard Gronke, Robert Boss, Brian Walls, Gelys Trancho, Inseok Song, Chris Carter, Peter Groskowski, Tatiana Paz, John White and James Patao.  We also thank our anonymous referee who provided many valuable comments which greatly improved the clarity and content of this paper.  This study was supported by the Gemini Observatory, which is operated by the Association of Universities for Research in Astronomy, Inc., on behalf of the international Gemini partnership of Argentina, Australia, Brazil, Canada, Chile, the United Kingdom, and the United States of America.

\clearpage



\clearpage 






 
\clearpage

\begin{deluxetable}{lcccccccc}
\tabletypesize{\scriptsize}
\tablecaption{Observing Log \label{tbl-1}}
\tablewidth{0pt}
\tablehead{
\colhead{Star} & \colhead{HH \#} & \colhead{Obs. Date} & \colhead{PA of} & 
\colhead {Exp Time/Coadds} & \colhead{\# Exp} & \colhead{Total Exposure} & 
\colhead{Note on} \\
\colhead{Name} & \colhead{} & \colhead{(UT)}  & \colhead{Obs.}   & 
\colhead {} & \colhead{} & \colhead{Time} & \colhead{Obs.} } 
\startdata
T Tau & HH 255 & 2005 Oct 25 & 0$^{\circ}$ &  5.3s/24 & 36 & 4580s & AO Flexure Test \\
RW Aur  & HH 229 & 2005 Oct 22 &   221$^{\circ}$ & 40s/1 & 11 & 440s & AO Guide Test \\
XZ Tau & HH 152 & 2005 Oct 25 &   0$^{\circ}$ & 30s/1 & 28 & 820s & AO+OIWFS$^*$ Guide Test \\
DG Tau & HH 158 & 2005 Oct 26 &   0$^{\circ}$ & 20s/6 & 101 & 12120s & AO+OIWFS Flexure Test \\ 
HV Tau C & HH 233 & 2005 Oct 22  &  114$^{\circ}$ & 900s/1 & 3 & 2700s & System Sensitivity Test \\
HL Tau & HH 150 & 2006 Feb 12 &   146$^{\circ}$ & 900s/1 & 3 & 2700s & $0.''2$ Occulting Disk SV \\
\enddata


\tablenotetext{*}{The NIFS On-Instrument WaveFront Sensor (OIWFS) is used to correct spatial flexure in the observations.}
\end{deluxetable}

\clearpage

\begin{deluxetable}{ccccccc}
\tabletypesize{\scriptsize}
\tablecaption{Percent H$_2$ Flux Coincident with Continuum Emission\label{tbl-2_01}}
\tablewidth{0pt}
\tablehead{
\colhead{   } & \colhead{T Tau}   & \colhead {RW Aur} & \colhead{XZ Tau} & \colhead{DG Tau} & \colhead{HV Tau C} & \colhead{HL Tau} }
\startdata
Percent H$_2$ {\it v}=1-0 S(1) & 3$^{+5}_{-3}$\% & 10$\pm$4\% & 23$\pm$6\% & 28$\pm$4\% & 59$\pm$6\% & 46$\pm$13 \\

\enddata

\end{deluxetable}

\clearpage

\begin{deluxetable}{lccccccccc}
\tabletypesize{\scriptsize}
\rotate
\tablecaption{H$_2$ Line Fluxes and 3$\sigma$ Detection Limits$^1$\label{tbl-3}}
\tablewidth{0pt}
\tablehead{
\colhead{Star} & \colhead{Extraction}  & \colhead {1-0 S(2)} & \colhead{2-1 S(3) } & \colhead{2-1 S(2)}& \colhead{1-0 S(0)}& \colhead{2-1 S(1)}& \colhead{1-0 Q(1)$^2$} & \colhead{1-0 Q(2)$^2$}& \colhead{1-0 Q(3)$^2$} \\
\colhead{   } & \colhead{Aperture Size}   & \colhead{2.03} & \colhead{2.07} & \colhead{2.15}& \colhead{2.22}& \colhead{2.24}& \colhead{2.40}& \colhead{2.41}& \colhead{2.42} }
\startdata
\multicolumn{10}{c}{Centered on Continuum Emission (Blue in Figure 3)} \\
T Tau & 1.$''$2 & 0.13$\pm$0.04 & $<$0.15 & $<$0.12 & 0.12$\pm$0.03 & $<$0.12 & 0.91$\pm$0.18 & 0.12$^{0.20}_{-0.12}$ & 1.07$\pm$0.22  \\
RWAur  & 1.$''$2 & $<$0.20 & $<$0.23 & $<$0.16 & $<$0.13 & $<$0.14 & $<$0.5 & $<$0.5 & $<$0.5 \\
XZ Tau & 0.$''$6 & 0.30$\pm$0.05 & $<$0.15 & $<$0.24 & 0.31$\pm$0.15 & $<$0.40 & 0.90$\pm$0.20 & 0.15$^{+0.24}_{-0.15}$ & 1.11$\pm$0.26 \\
DG Tau & 0.$''$8 & 0.23$\pm$0.06 & $<$0.20 & $<$0.24 & 0.21$\pm$0.09 & $<$0.24 & 1.11$\pm$0.31 & 0.16$^{+0.22}_{-0.16}$ & 1.24$\pm$0.38  \\ 
HV Tau C & 1.$''$2 &  0.343$\pm$0.006 & 0.073$\pm$0.07 & 0.026$\pm$0.008 & 0.226$\pm$0.006 & 0.080$\pm$0.007 & 1.090$\pm$0.006 & 0.302$\pm$0.006 & 1.054$\pm$0.005 \\
HL Tau & 1.$''$2 &  $<$0.21 & $<$0.24 & $<$0.22 & $<$0.19 & $<$0.19 & $<$0.5 & $<$0.5 & $<$0.5 \\
\hline
\multicolumn{10}{c}{Centered on Peak H$_2$ Emission and Knots (Green in Figure 3)} \\
T Tau & 0.$''$2 & 0.28$\pm$0.02 & 0.07$\pm$0.03 & 0.04$\pm$0.02 & 0.20$\pm$0.01 & 0.06$\pm$0.01 & 0.87$\pm$0.09 & 0.21$\pm$0.05 & 0.82$\pm$0.09  \\
RWAur  & 0.$''$2  &  0.25$\pm$0.04 & 0.12$\pm$0.05 & $<$0.09 & 0.17$\pm$0.04 & 0.11$\pm$0.04 & 0.76$\pm$0.16 & 0.25$\pm$0.18 & 0.73$\pm$0.20  \\
XZ Tau & 0.$''$2 &  0.32$\pm$0.02 & 0.12$\pm$0.04 & 0.05$\pm$0.02 & 0.19$\pm$0.004 & 0.11$\pm$0.04 & 0.80$\pm$0.14 & 0.31$\pm$0.15 & 0.85$\pm$0.17  \\
DG Tau & 0.$''$2  &  0.24$\pm$0.02 & $<$0.06 & $<$0.06 & 0.28$\pm$0.03 & 0.07$\pm$0.03 & 1.00$\pm$0.24 & 0.26$\pm$0.26 & 1.03$\pm$0.28  \\ 
HL Tau & 0.$''$2 &  0.21$\pm$0.04 & 0.09$\pm$0.04 & $<$0.09 & 0.31$\pm$0.04 & 0.08$\pm$0.04 & 0.26$\pm$0.21 & 0.22$\pm$0.22 & 0.42$\pm$0.23  \\

\enddata


\tablenotetext{1}{In all cases, the H$_2$ line fluxes are normalized by the {\it v}=1-0 S(1) line at 2.12$\mu$m. The  {\it v}=1-0 S(1) line is not included in this table because all values are 1.0.}
\end{deluxetable}

\clearpage

\begin{deluxetable}{lccc}
\tabletypesize{\scriptsize}
\tablecaption{A$_v$ and T$_{ex}$ from H$_2$ Ratios\label{tbl-2_02}}
\tablewidth{0pt}
\tablehead{
\colhead{Star Name} & \colhead{A$_v$ (1-0 Q(3)/S(1) ratio)}  & \colhead{A$_v$ (1-0 Q(2)/S(0) ratio)}   & \colhead {T$_{ex}$ (K)$^1$} }
\startdata
T Tau &  8.0($\pm$5) & $<$0($\pm$12) & 1990$\pm$50 \\
RWAur  & 2.3($\pm$9) & 10($\pm$20) & 2260$\pm$60 \\
XZ Tau & 9.7($\pm$9) & 14($\pm$15) & 2270$\pm$50 \\
DG Tau & 11.2($\pm$10) & $<$0($\pm$19) & 1810$\pm$60 \\ 
HV Tau C & 20.3($\pm$0.5) & 7($\pm$1) & 1920$\pm$40 \\
HL Tau & $<$0.0 & $<$0.0 & 2030$\pm$90 \\
\enddata
\tablecomments{Visual extinction values derived from H$_2$ line ratios provide approximations of extinctions toward the H$_2$ emitting region assuming an ISM extinction law for the obscuring material.  Uncertainties in extinction values can be large, and are derived from the uncertainties in the H$_2$ emission line ratios.  Visual extinctions of less than zero for HL Tau likely arise from strong scattered light within the IFU field and from poor telluric correction in the Q-branch region.  Excitation Temperatures were derived by dereddening line fluxes and determining a median temperature from the linear fit to the LTE level populations (Figure 4).}

\end{deluxetable}

\clearpage

\begin{deluxetable}{lccc}
\tabletypesize{\scriptsize}
\tablecaption{H$_2$ {\it v=1-0} S(1) Velocities and Dispersions \label{tbl-2_03}}
\tablewidth{0pt}
\tablehead{
\colhead{Star Name} & \colhead{v$_r$(system)}   & \colhead {H$_2$ v$_c$} & \colhead{H$_2$ $\sigma_v$} }
\startdata
Sky Line & -- & 0.3$\pm$2.6km/s & 30.2$\pm$2.5km/s \\
\hline
T Tau & 20.2km/s & -14.1$\pm$24.2km/s & 30.0$\pm$23.8km/s  \\
RWAur  & 16.0km/s & 43.7$\pm$28.5km/s & 55.0$\pm$50.3km/s \\
XZ Tau & 18.0km/s & -7.3$\pm$5.1km/s & 27.3$\pm$3.3km/s \\
DG Tau & 19.3km/s & -2.4$\pm$18.4km/s & 32.2$\pm$18.9km/s \\ 
HV Tau C & 23.1km/s & -10.1$\pm$11.7km/s & 28.4$\pm$12.9km/s \\
HL Tau & 18.3km/s & -8.9$\pm$4.1km/s & 26.5$\pm$31.5km/s \\
\enddata
\tablecomments{Emission line central velocities and line dispersions were calculated for the {\it v}=1-0 S(1) emission in every IFU spectral pixel than had an emission S/N greater than $\sim$12. H$_2$ velocities are corrected for systemic radial velocity and heliocentric effects.  Median values and uncertainties for v$_c$ and $\sigma_v$ were determined from the multiple velocity measurements over all spectral pixels.  Values for the sky emission line are included to demonstrate the average uncertainty measurements for the instrument.}
\end{deluxetable}

\clearpage

\begin{deluxetable}{l|c|cccccc|cc}
\tabletypesize{\scriptsize}
\tablecaption{Line Ratio Excitation}
\tablewidth{0pt}
\tablehead{
\colhead{Line Ratio} & \colhead{UV}  & \colhead {T} & \colhead{RW} & \colhead{XZ} & \colhead{DG} & \colhead{HL} & \colhead{HV} & \colhead{C Type} & \colhead{J Type} \\
\colhead{  } & \colhead{Pumped}  & \colhead {Tau} & \colhead{Aur} & \colhead{Tau} & \colhead{Tau} & \colhead{Tau} & \colhead{Tau C} & \colhead{Shocks} & \colhead{Shocks} }
\startdata
2-1 S(1)/1-0 S(1) & 0.55 & 0.06 & 0.11 & 0.10 & 0.06 & 0.10 & 0.07 & 0.05 & 0.24 \\ 
1-0 S(2)/1-0 S(0) & 1.09 & 1.58 & 1.52 & 2.00 & 1.15 & 0.67 & 2.07 & 1.56 & 2.08 \\
2-1 S(1)/2-1 S(3) & 1.58 & 0.78 & 1.09 & 1.20 & --  & 3.44 & 1.18 &  1.08 & 7.21 \\
1-0 S(1)/1-0 Q(1) & 1.00 & 1.38 & 1.38 & 1.51 & 1.47 & 3.48 & 1.35 & 1.29 & 1.59 \\
\hline

\enddata
\tablecomments{Calculations of the UV pumped fluorescent H$_2$ line ratios are from Black \& van Dishoeck (1987), line ratios arising from C and J type shocks are from Smith (1995).}
\end{deluxetable}

\clearpage

\begin{figure}
\epsscale{0.7}
\includegraphics[angle=0,width=17.0cm,height=12.5cm]{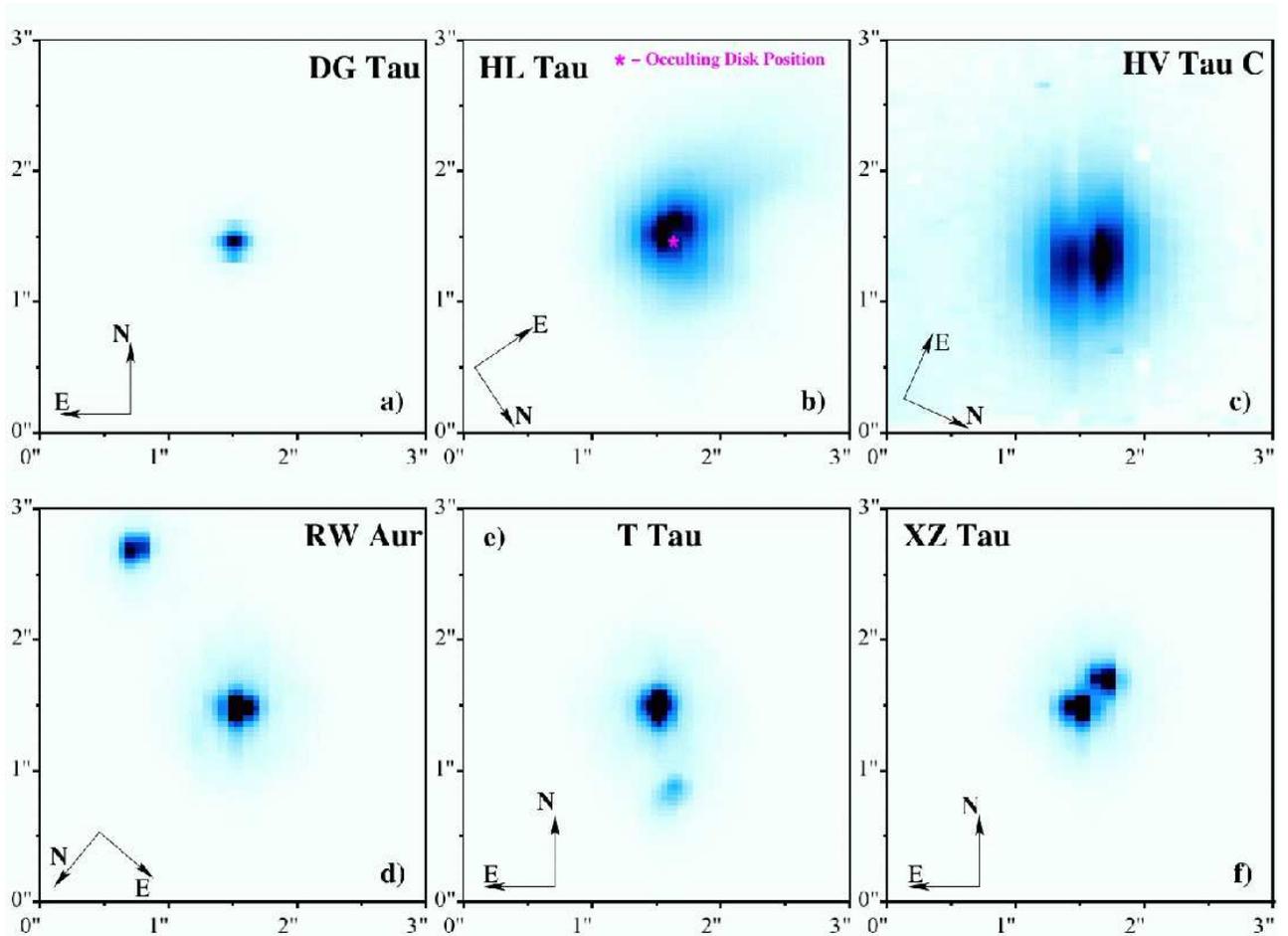}
\caption{Maps of the continuum emission at 2.12$\mu$m determined by fitting a line to the continuum emission blueward and redward of the {\it v}=1=0 S(1) line.  Images are scaled from +10\% to +90\% of the peak flux (5\% to 90\% for T Tau to display the position of the southern companions).}
\label{fig1}
\end{figure}

\begin{figure}
\epsscale{0.7}
\includegraphics[angle=0,width=17.0cm,height=12.5cm]{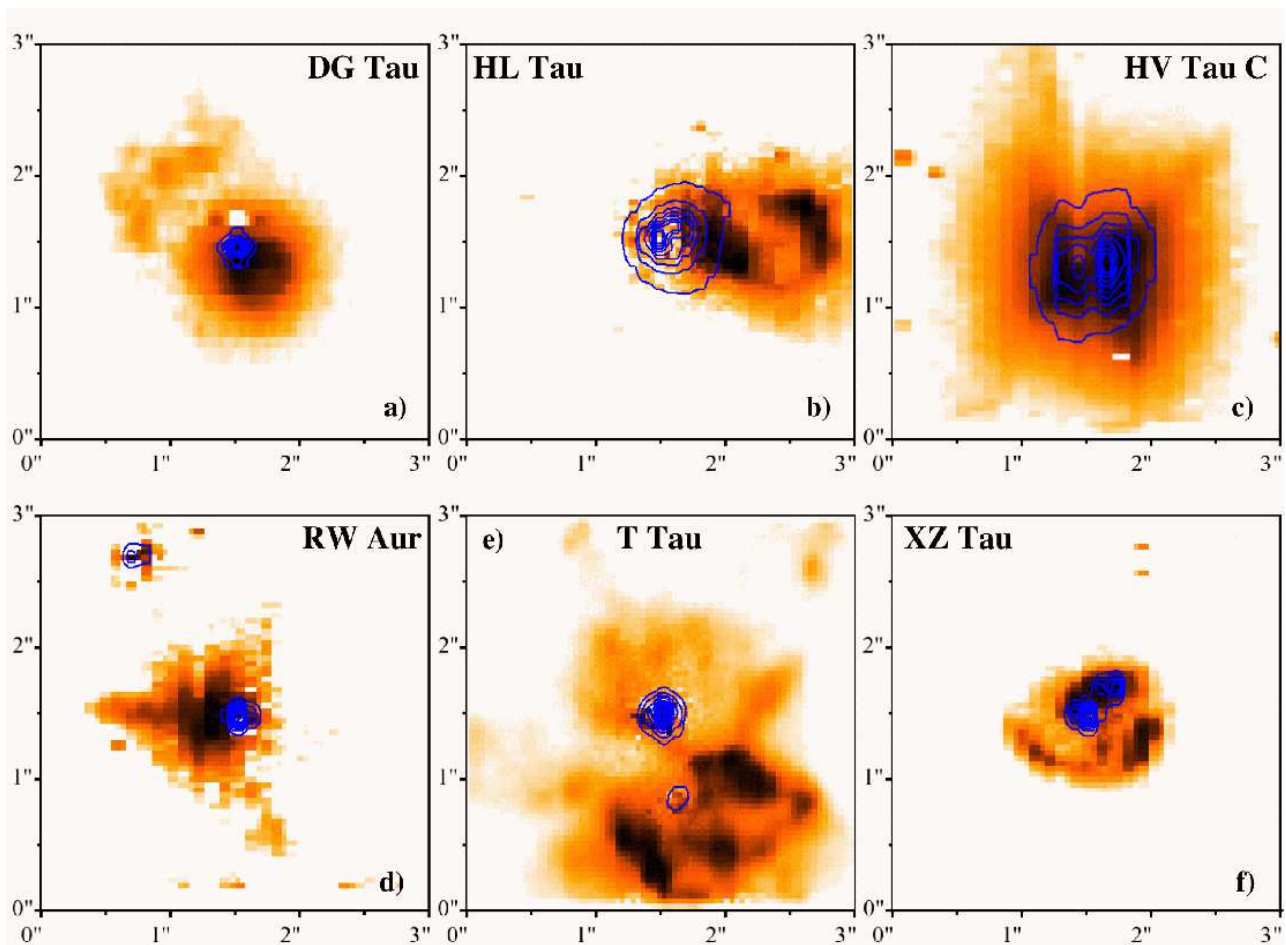}
\caption{Maps of the spatially extended {\it v}=1=0 S(1) H$_2$ emission for each star plotted as signal-to-noise (effectively a square root map of the detected flux at spatially extended distances from the star, to enhance lower level emission).  Overplotted in blue are nine contours of the continuum emission from +10\% to +90\% of the peak, T Tau has an additional contour at +5\% of the peak to identify the position of T Tau S.  For all except HV Tau C, the majority of the H$_2$ is not spatially coincident with the continuum emission. Observations of HL Tau were acquired with an 0,$''2$ diameter occulting disk blocking the bright flux from the central star (Figure 1).  The orientation in the figure is consistent with that presented in Figure 1.}
\label{fig2}
\end{figure}

\clearpage
\begin{figure}
\centering
\includegraphics[angle=90, width=.45\textwidth]{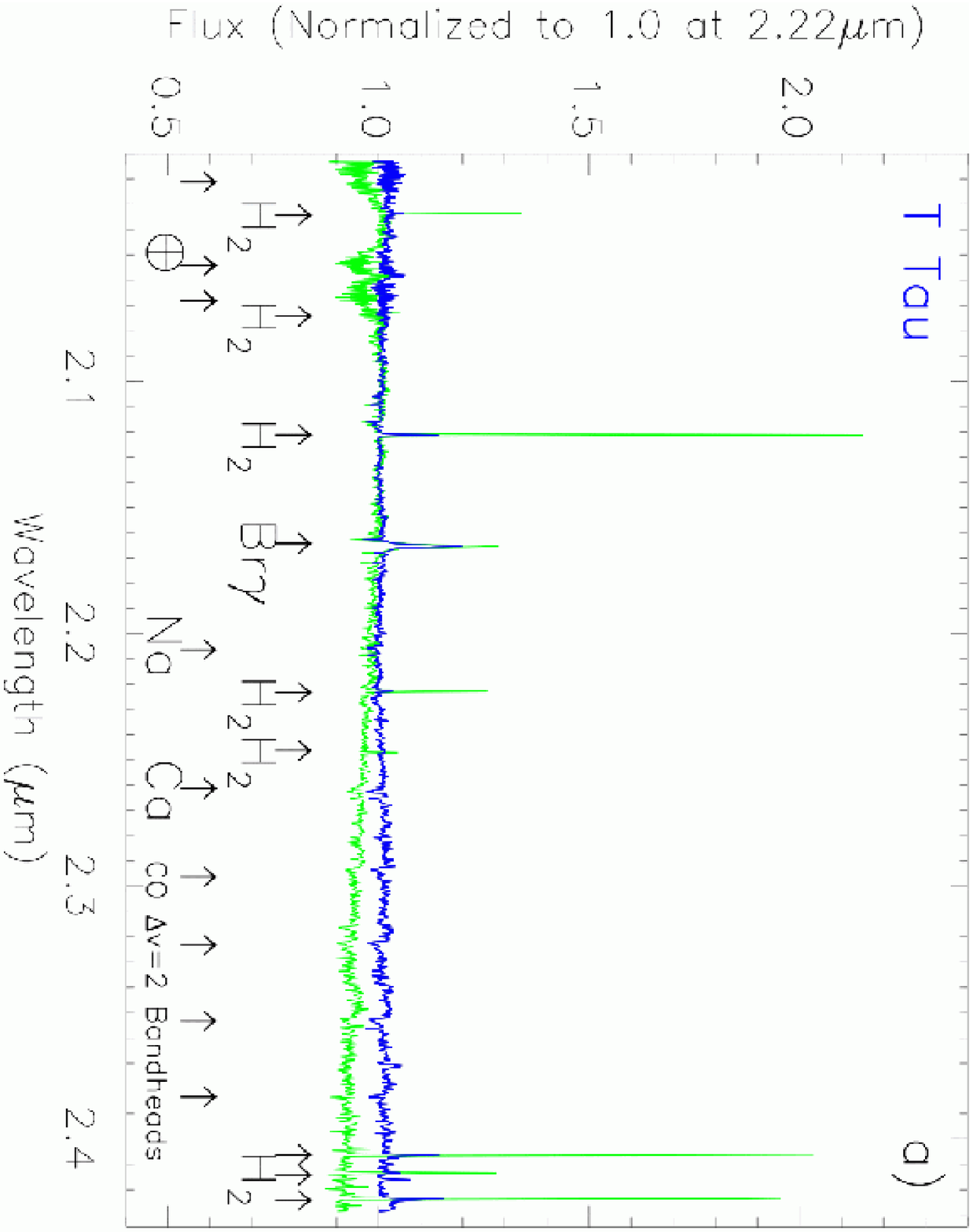}
\includegraphics[angle=90, width=.45\textwidth]{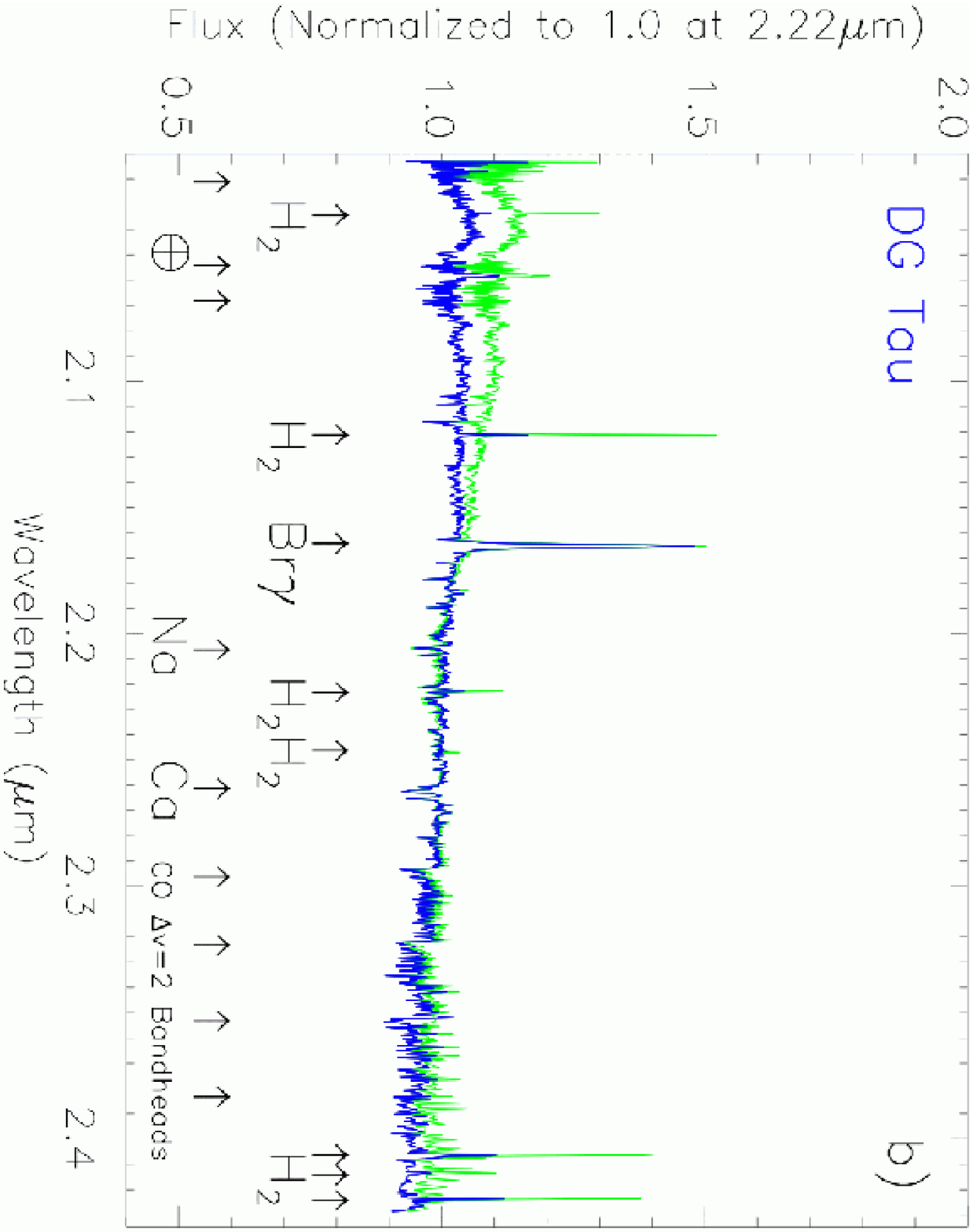}\\
\includegraphics[angle=90, width=.45\textwidth]{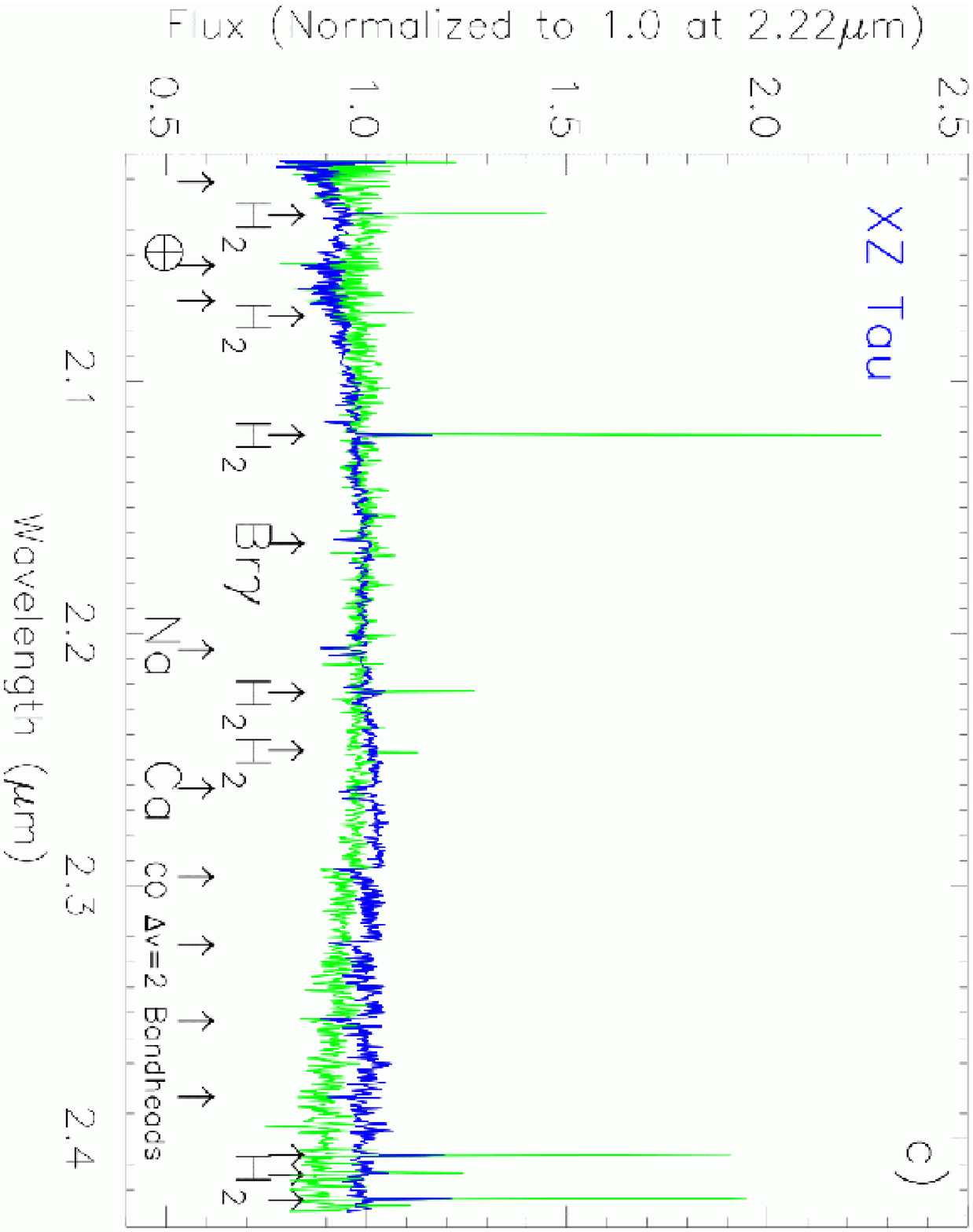}
\includegraphics[angle=90, width=.45\textwidth]{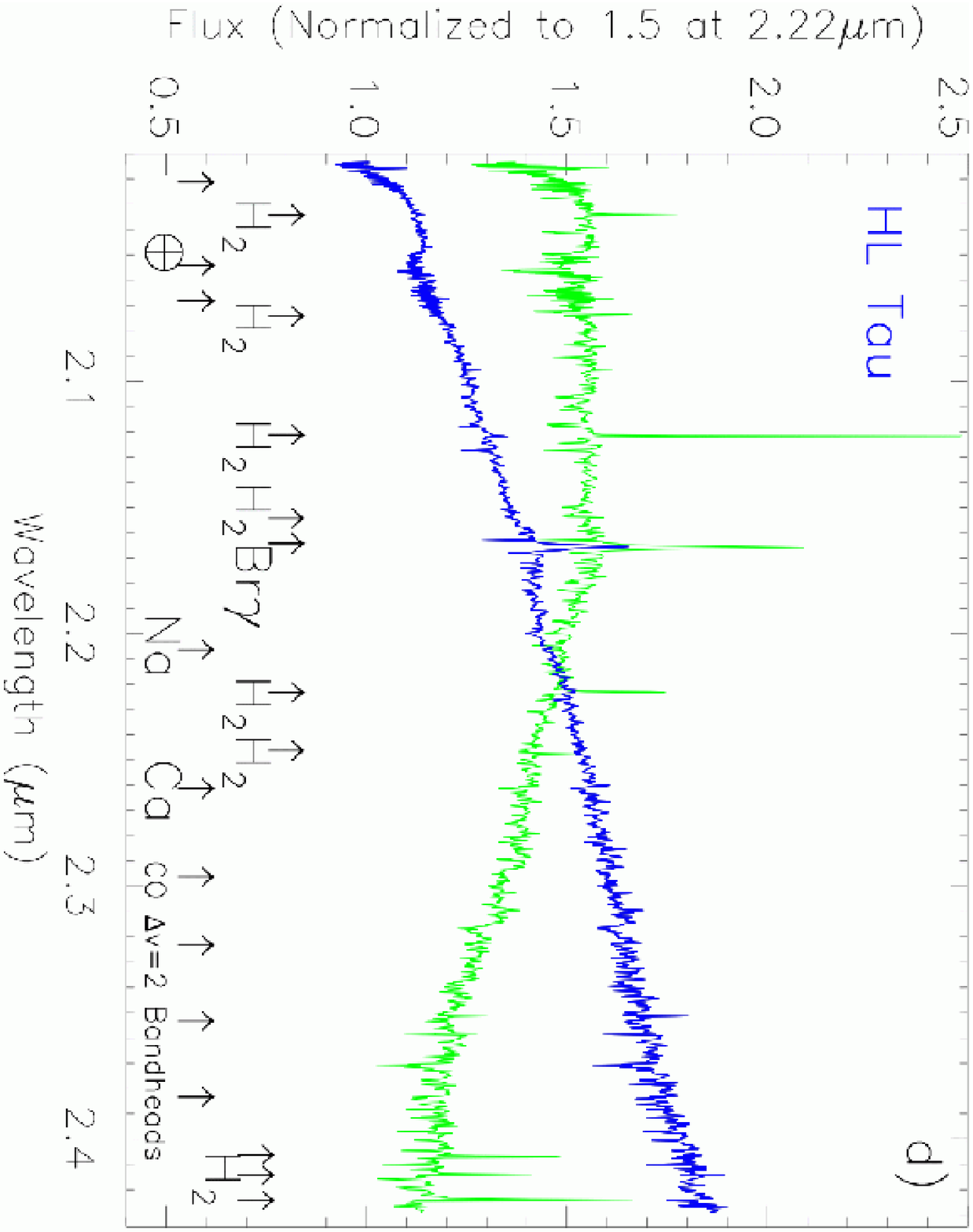}\\
\includegraphics[angle=90, width=.45\textwidth]{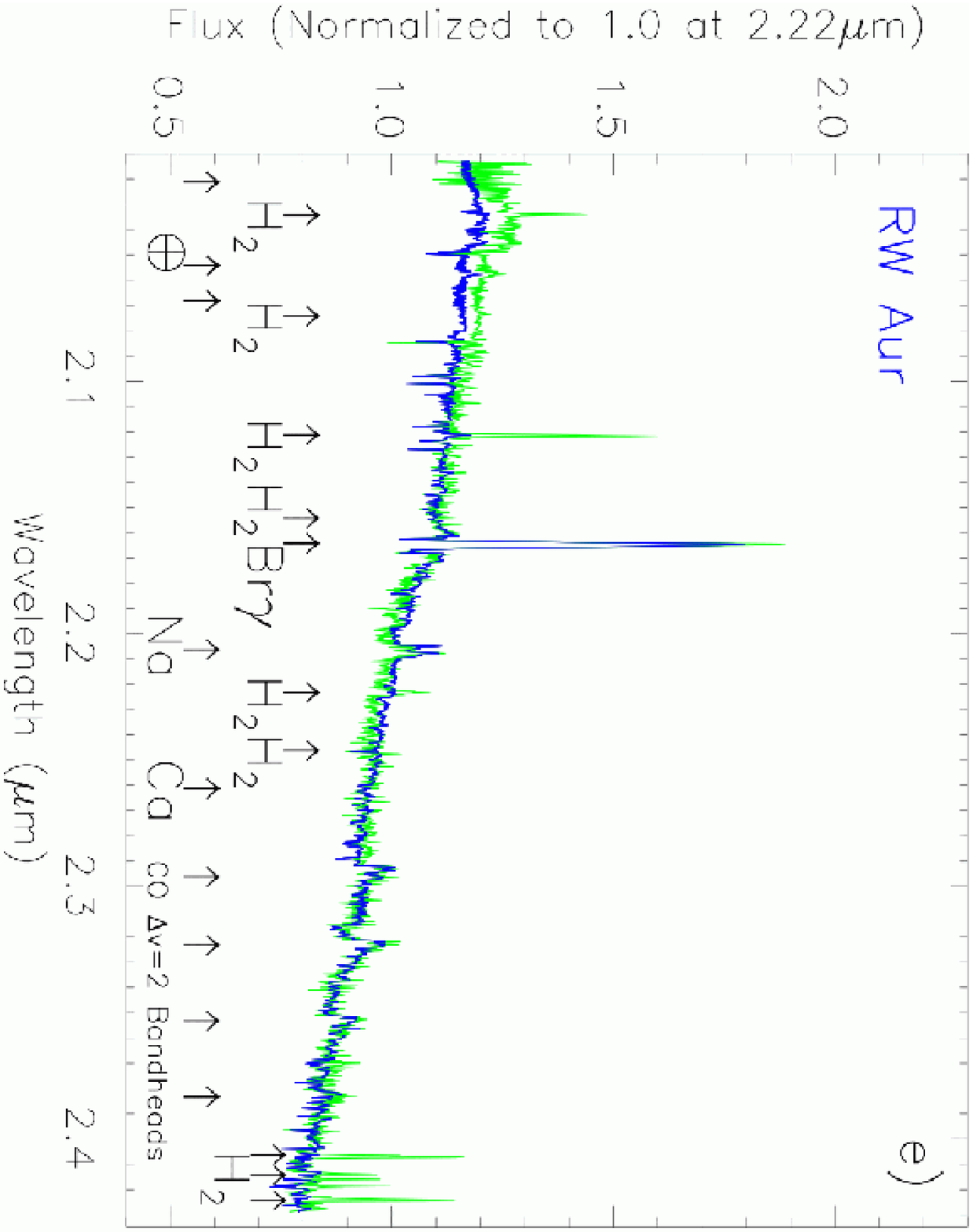}
\includegraphics[angle=90, width=.45\textwidth]{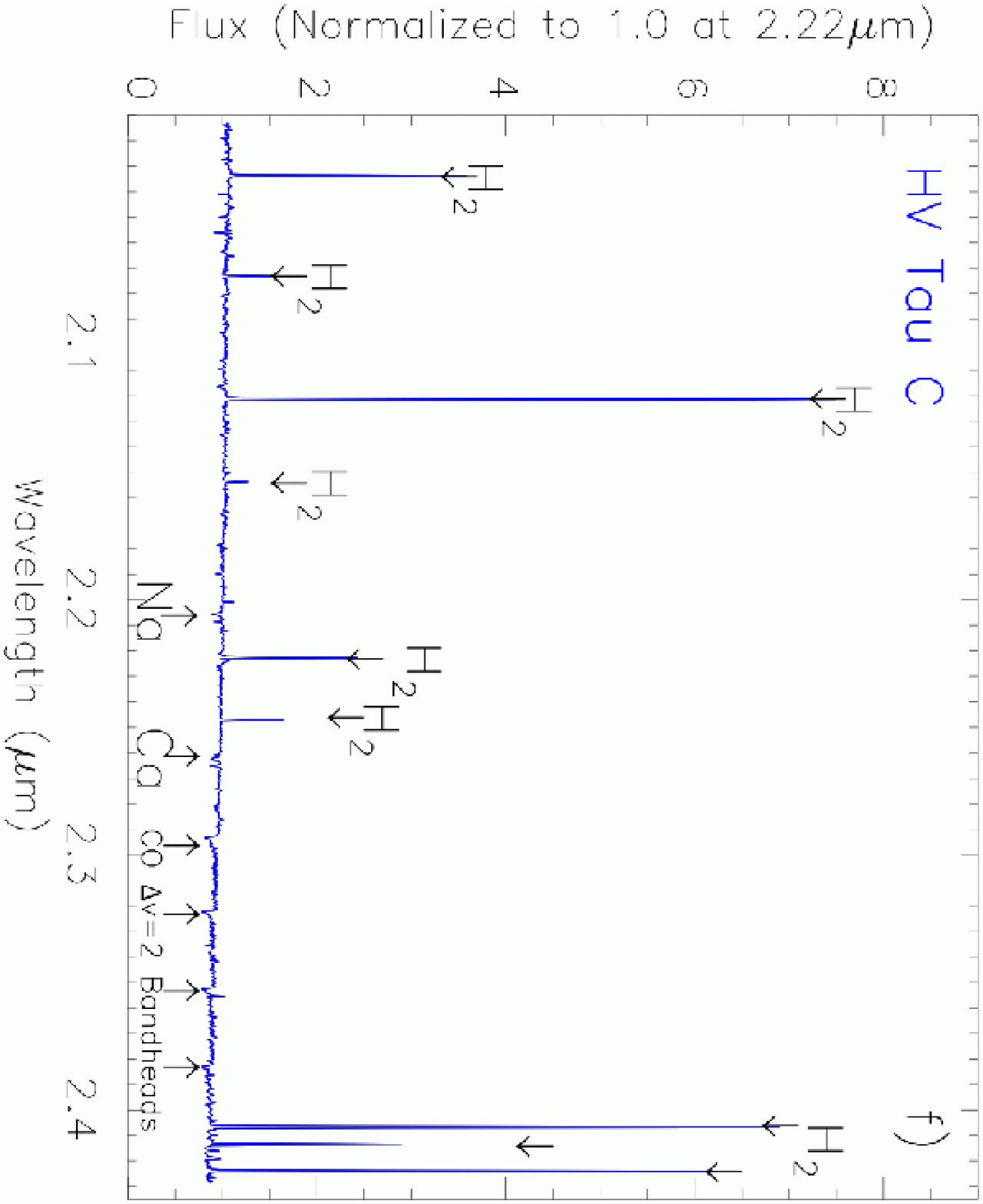}
\caption{The 2.0-2.4$\mu$m spectra of the six Classical T Tauri Stars observed for this study.  The blue spectrum is the data extracted using a wide aperture and centered on the peak continuum emission (see Table 3).  To enhance the detectability of the H$_2$ features over the continuum emission, a spectrum was also extracted using an 0.$''$2 radius aperture centered at the location of the peak of the spatially extended H$_2$ emission (plotted in green). \label{fig3}}
\end{figure}
\clearpage

\begin{figure}
\epsscale{0.7}
\includegraphics[angle=90,width=17.0cm,height=13.0cm]{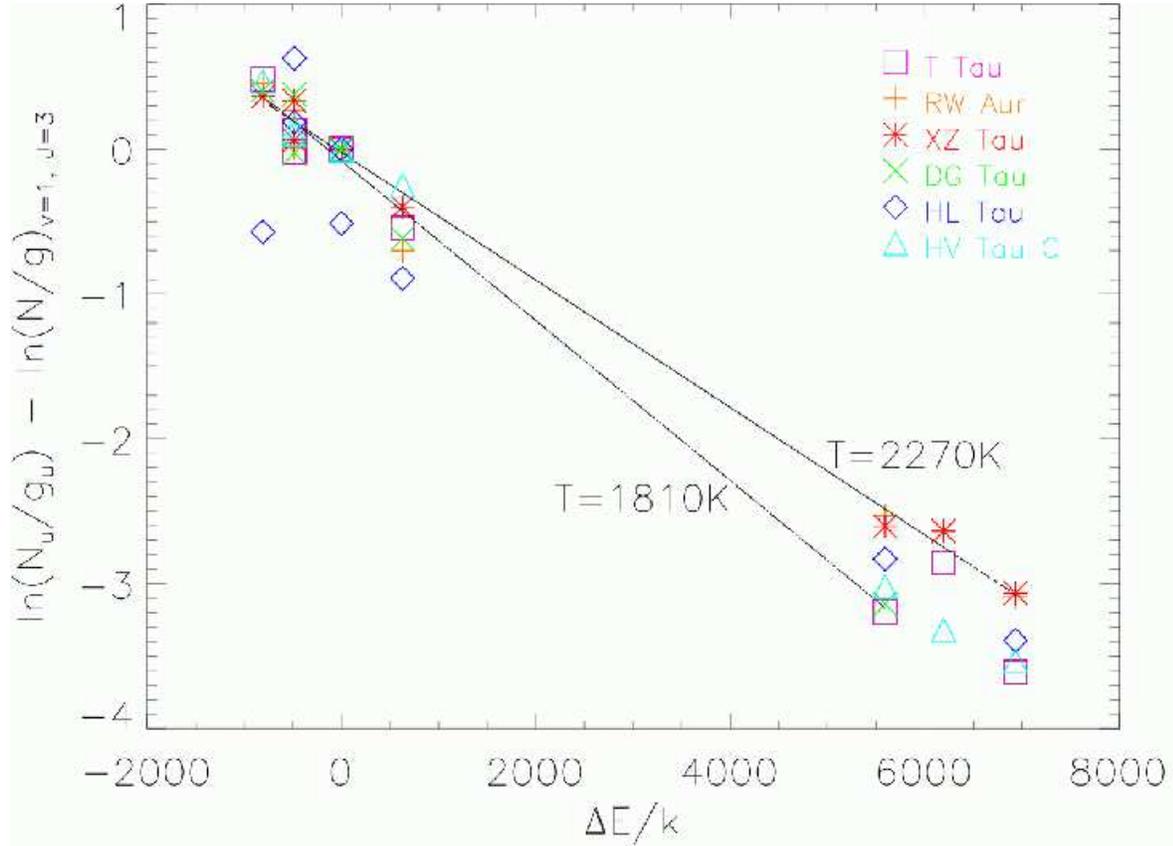}
\caption{LTE Energy Population diagrams for the six stars derived from dereddened line fluxes for multiple H$_2$ transitions.  The best-fit linear slopes to the observed H$_2$ populations correspond to gas temperatures in the range of 1800-2300K (Table 4).  Lines with slopes corresponding to T=1810 and T=2270 are overplotted (fits to DG Tau and XZ Tau, respectively).}
\label{fig4}
\end{figure}

\begin{figure}
\epsscale{0.7}
\includegraphics[angle=0,width=17.0cm,height=7.0cm]{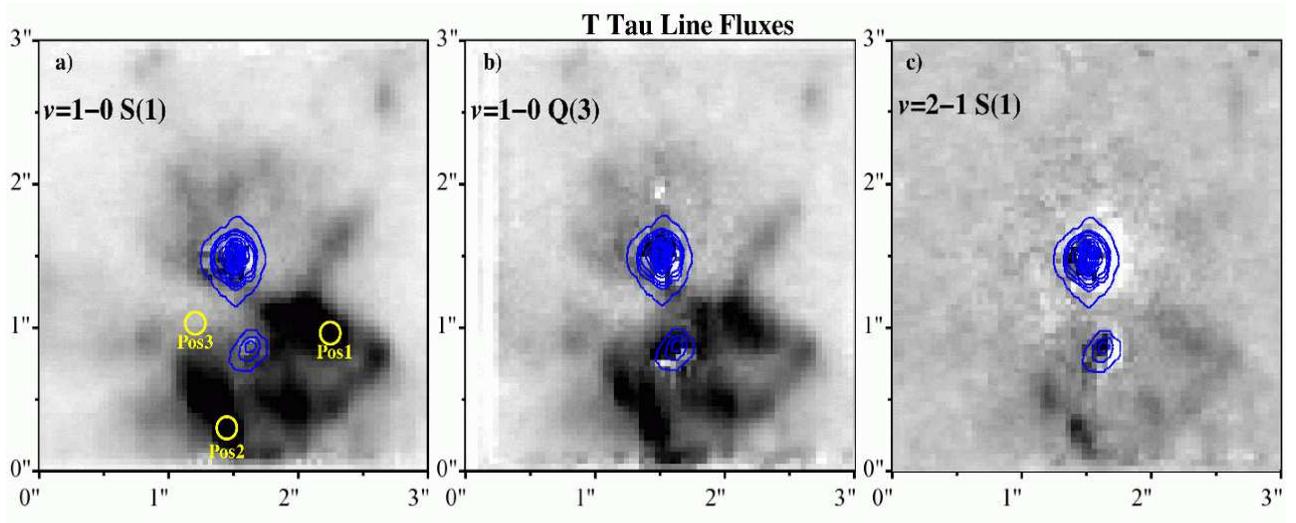}
\caption{Emission line maps in the vicinity of T Tau triple of three H$_2$ species:  {\it v}=1-0 S(1) (a), {\it v}=1-0 Q(3) (b) and {\it v}=2-1 S(1) (c). The images are displayed linearly in flux from -10\% to +90\% of the peak in the  {\it v}=1-0 S(1) map for a and b.  To enhance the low level emission structure in panel c (for the {\it v}=2-1 S(1) emission), the linear scaling of the image is from -10\% to +30\% of the peak flux in the {\it v}=1-0 S(1) map. Overplotted in panel a are the locations of three 0.$''$2 circular apertures where emission lines were extracted for comparison of ratios (See Figure 6).}
\label{fig5}
\end{figure}

\begin{figure}
\centering
\begin{tabular}{cc}
\includegraphics[angle=90,width=.45\textwidth]{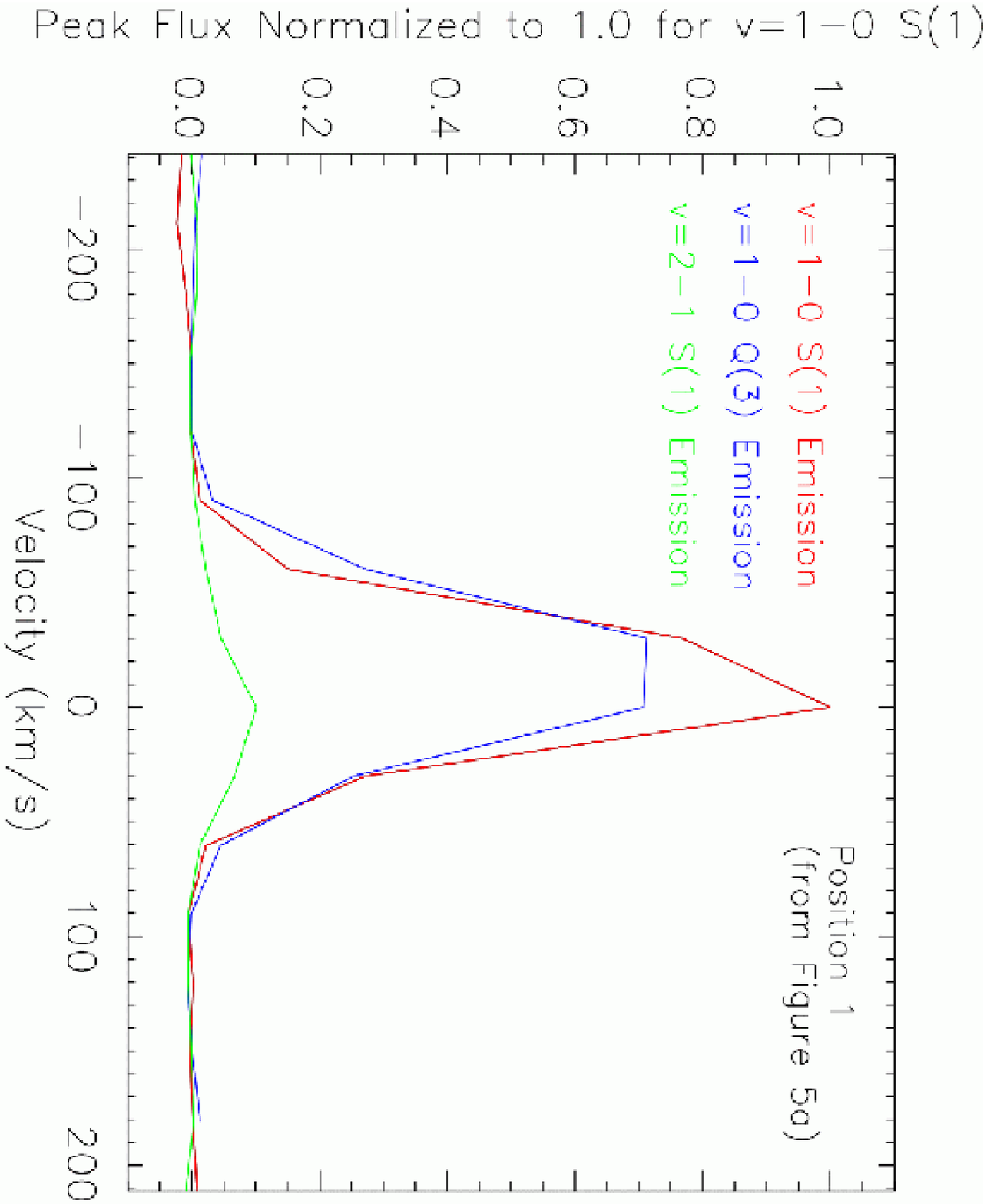} &
\includegraphics[angle=90,width=.45\textwidth]{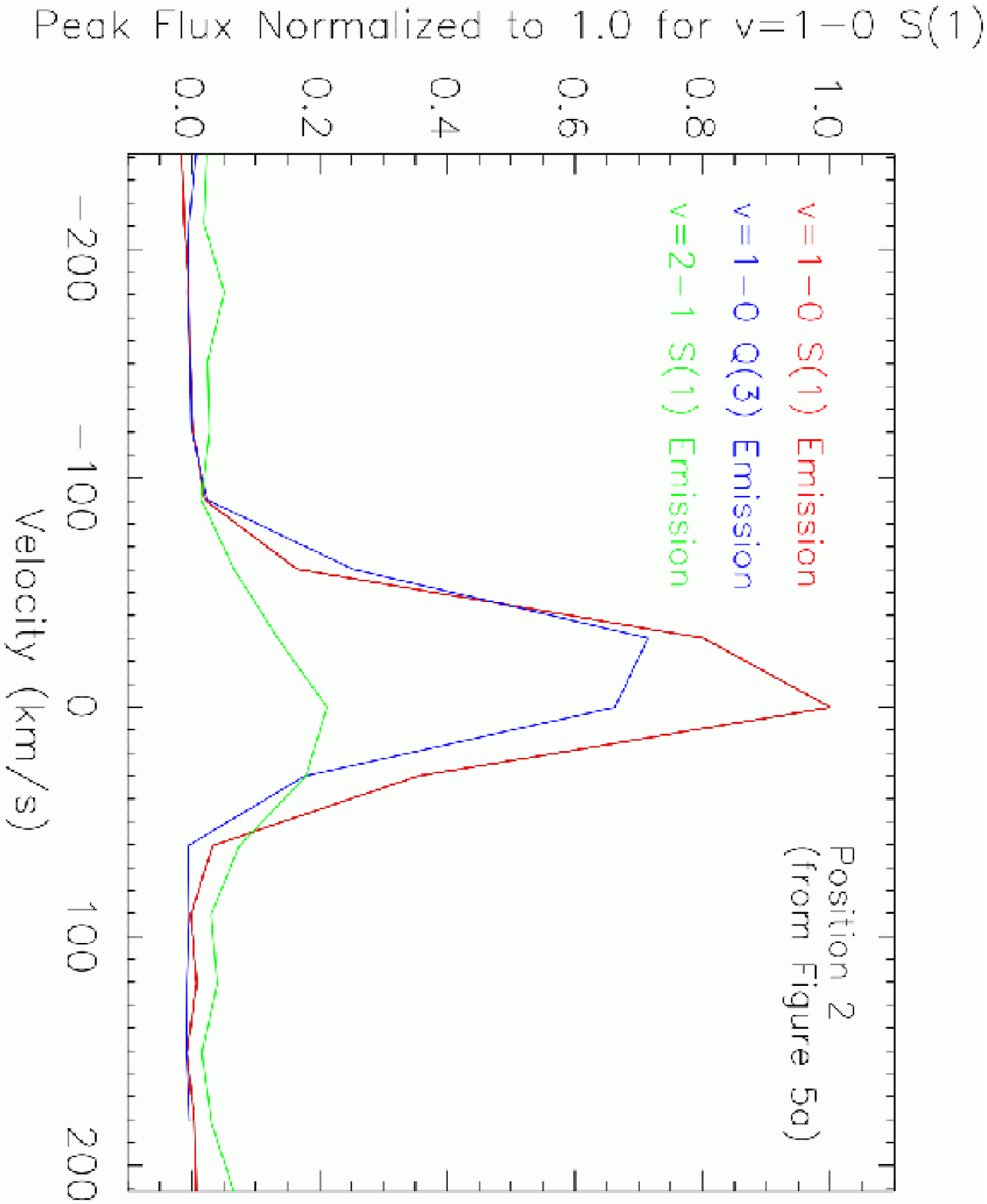} \\
\includegraphics[angle=90,width=.45\textwidth]{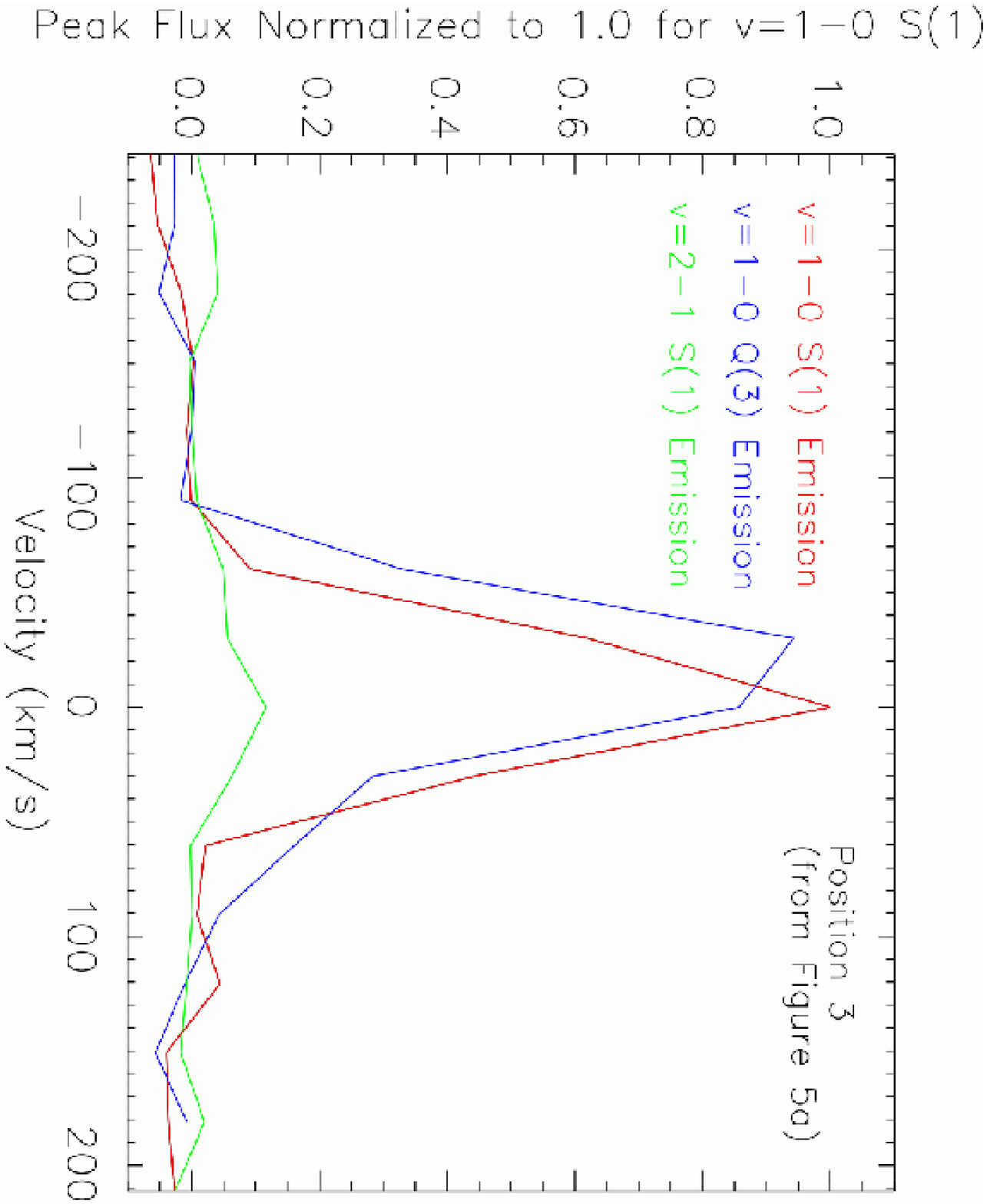}
\end{tabular}
\caption{Plots of the {\it v}=1-0 S(1) (red), {\it v}=1-0 Q(3) (blue) and {\it v}=2-1 S(1) (green) emission lines extracted in the vicinity of T Tau with 0.$''$2 circular apertures at each of three positions defined in the spatial map in Figure 5a.  At the three different positions, the line ratios of these three species vary on a significant level.  Variations in line ratio are caused by differences in the amount of obscuring material along the line of sight (1-0 Q(3) to S(1)) and by differences in the H$_2$ excitation temperature (2-1 S(1) to 1-0 S(1)).\label{fig6}}
\end{figure}

\begin{figure}
\epsscale{0.7}
\includegraphics[angle=0,width=17.0cm,height=10.2cm]{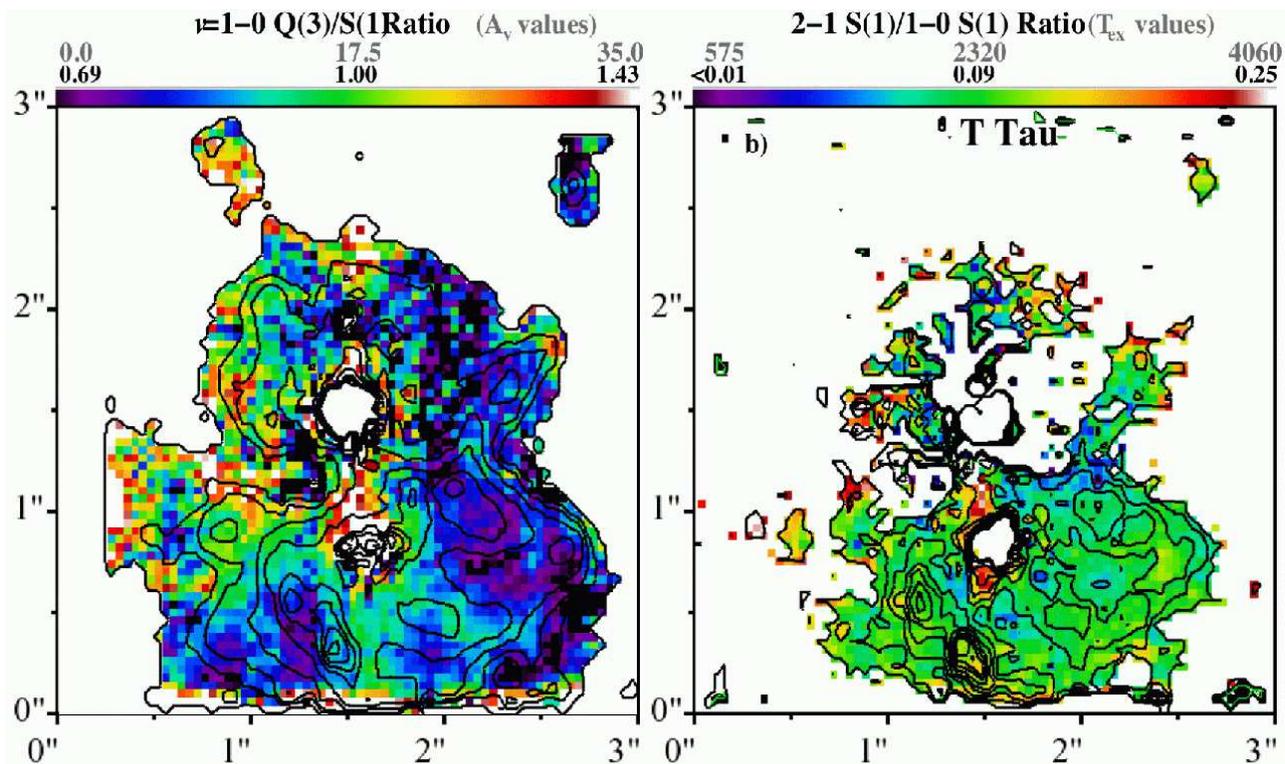}
\caption{Maps of the approximate visual extinction (A$_v$) and H$_2$ excitation temperature (T$_{ex}$) spatial structure derived from resolved {\it v}=1-0 Q(3)/S(1) and (dereddened) 2-1 S(1)/1-0 S(1) line ratios, respectively.  Overplotted contours are the signal-to-noise in the Q(3) and 2-1 S(1) emission.  The lowest contour is at a level of S/N=8.0 for Q(3) and S/N=15.0 for S(1), and the highest contour is at 90\% of the peak line flux.  The color scale represents {\it v}=1-0 Q(3)/S(1) line ratios of 0.0 to 1.4 (corresponding to approximate A$_v$ values of 0.0 to 35.0) and {\it v}=2-1 S(1)/{\it v}=1-0 S(1) ratios of 0.01 to 0.25 (tracing T$_{ex}$ from 575K to 4060K).  Values for the A$_v$ and T$_{ex}$ are included in gray in the figure key.  The apparent enhancement of extinction at the position of T Tau South (from Figure 1) is {\it not} genuine, investigation of the data shows that it arises instead from low intrinsic H$_2$ flux and imperfect subtraction of the strong continuum emission in the spectral region of the Q(3) transition.}
\label{fig7}
\end{figure}

\begin{figure}
\epsscale{0.7}
\includegraphics[angle=0,width=17.0cm,height=10.2cm]{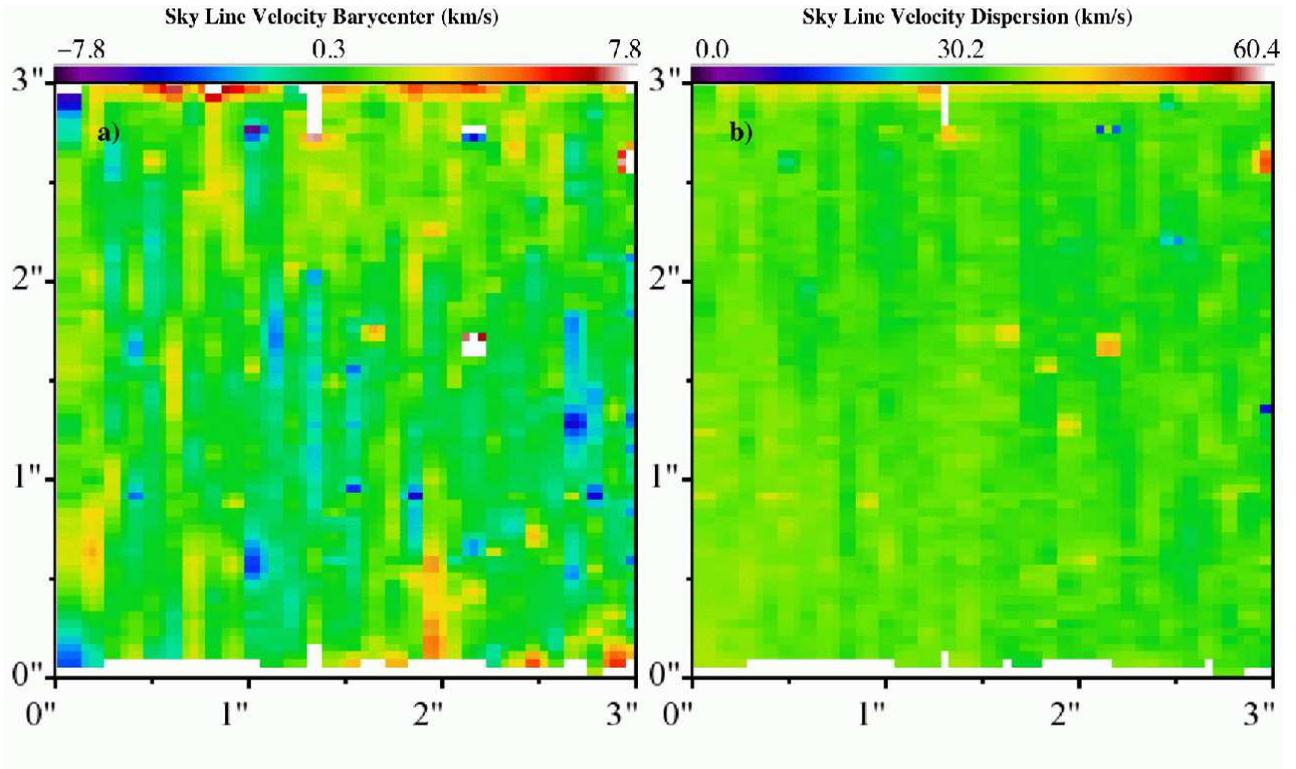}
\caption{Maps of the velocity barycenter (left) and dispersion (right) for the sky emission line.  Analysis of these values for the sky line provides a measure of the stability and accuracy of the v$_c$ and $\sigma_{v}$ calculations over the NIFS IFU field.}
\label{fig8}
\end{figure}

\begin{figure}
\epsscale{0.7}
\includegraphics[angle=0,width=17.0cm,height=12.5cm]{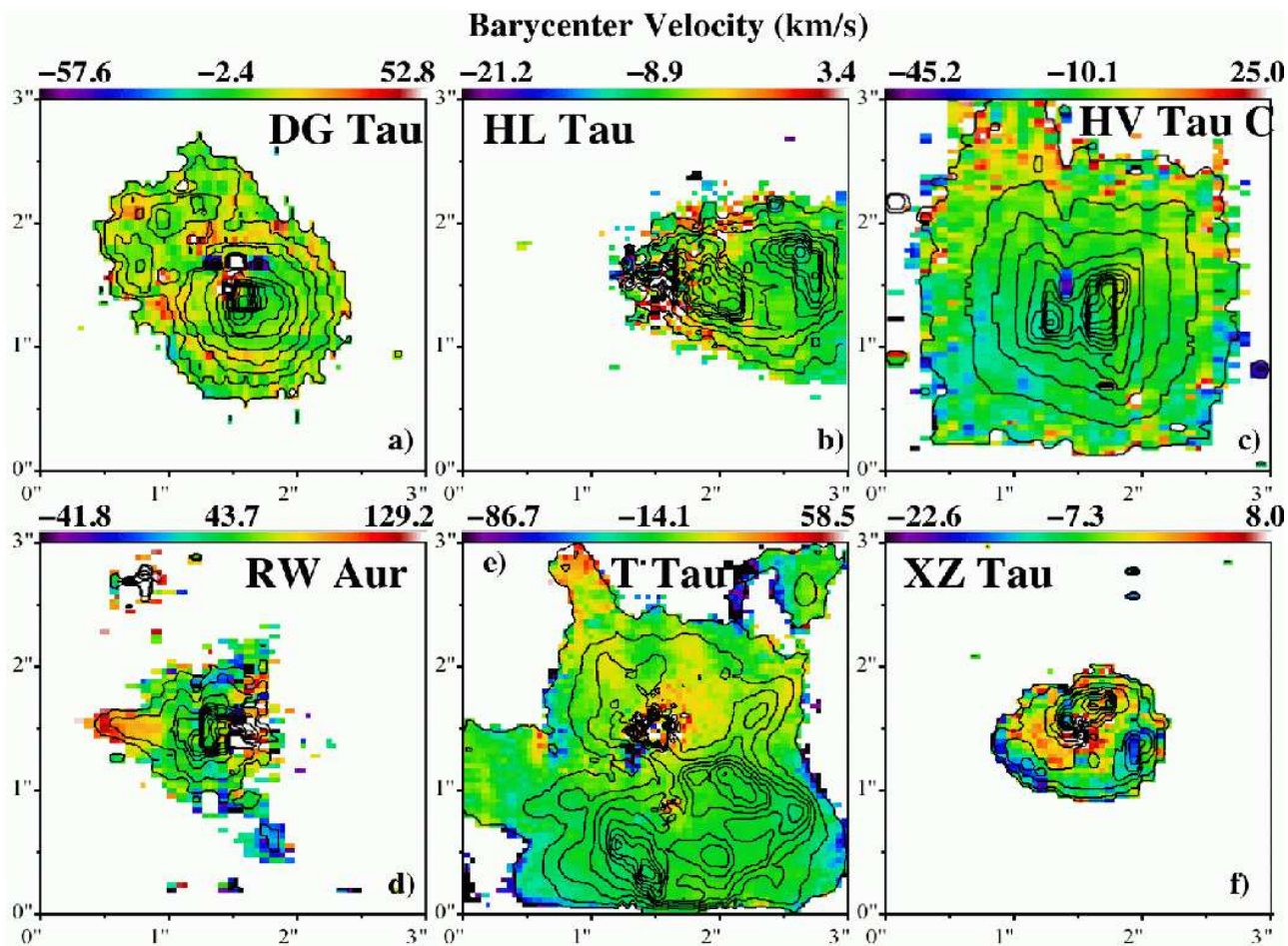}
\caption{Maps of the velocity barycenters, v$_c$, of the {\it v}=1=0 S(1) emission.  Each star is plotted with the median velocity in green and a color scale ranging from +3$\sigma$ to -3$\sigma$ from the median value (where $\sigma$ is the standard deviation of the multiple v$_c$ measurements and each spectral pixel; Table 4).  Overplotted in black are contours of the signal-to-noise of the H$_2$ emission, the lowest contour corresponds to a S/N$\sim$12 (RW Aur has a lowest contour of S/N=18 to limit noise.  Observations of HL Tau were acquired with an 0,$''2$ diameter occulting disk blocking the bright flux from the central star (Figure 1).  The orientation in the figure is consistent with that presented in Figure 1.\label{fig9}}
\end{figure}

\begin{figure}
\epsscale{0.7}
\includegraphics[angle=0,width=17.0cm,height=12.5cm]{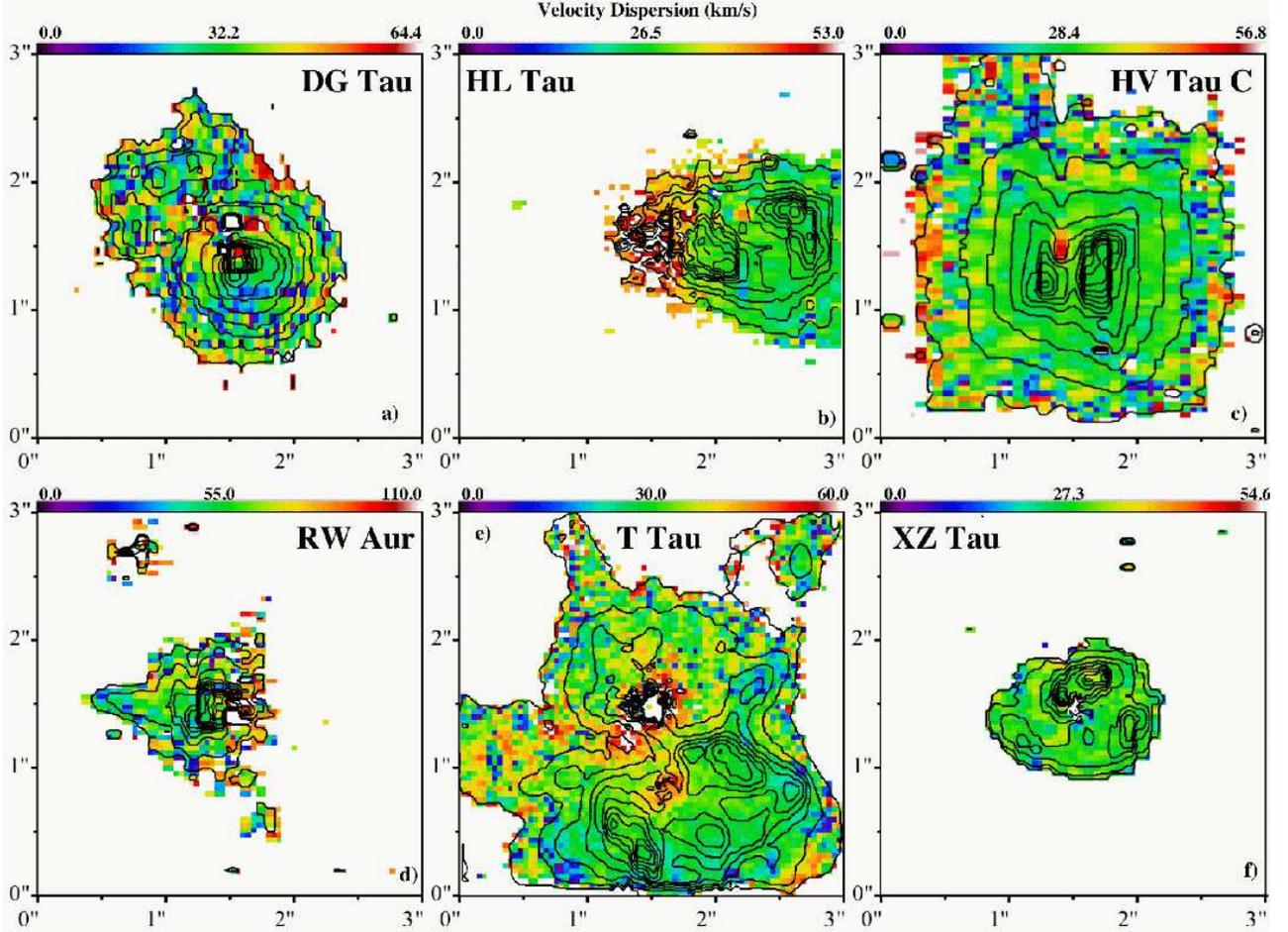}
\caption{Maps of the velocity dispersions, $\sigma_v$, of the {\it v}=1=0 S(1) emission.  Green represents the median dispersion for each star, and the color scale ranges from 0 km/s to twice the median dispersion.  Overplotted in black are contours of the signal-to-noise of the H$_2$ emission, as in Figure 8.   Observations of HL Tau were acquired with an 0,$''2$ diameter occulting disk blocking the bright flux from the central star (Figure 1).  The orientation in the figure is consistent with that presented in Figure 1.}
\label{fig10}
\end{figure}

\begin{figure}
\epsscale{0.7}
\includegraphics[angle=0,width=17.0cm,height=6.5cm]{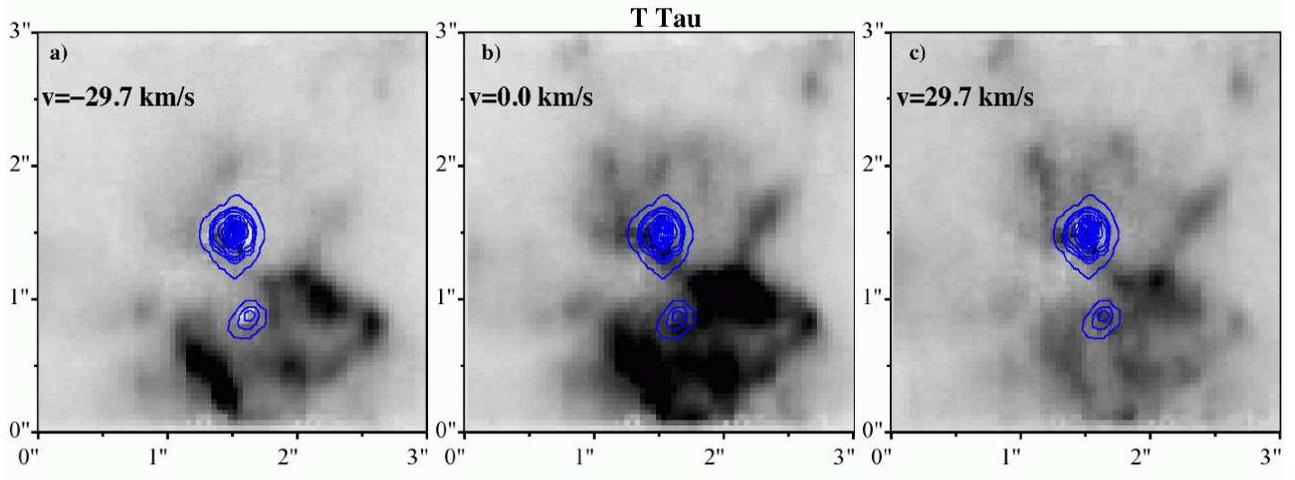}
\caption{Velocity channel map of spatially extended H$_2$ emission in the vicinity of T Tau.  The three channel maps show H$_2$ from -29.7km/s to +29.7km/s, and each of the three panels represents one velocity slice through the H$_2$ datacube.}
\label{fig11}
\end{figure}

\begin{figure}
\epsscale{0.7}
\includegraphics[angle=0,width=17.0cm,height=6.5cm]{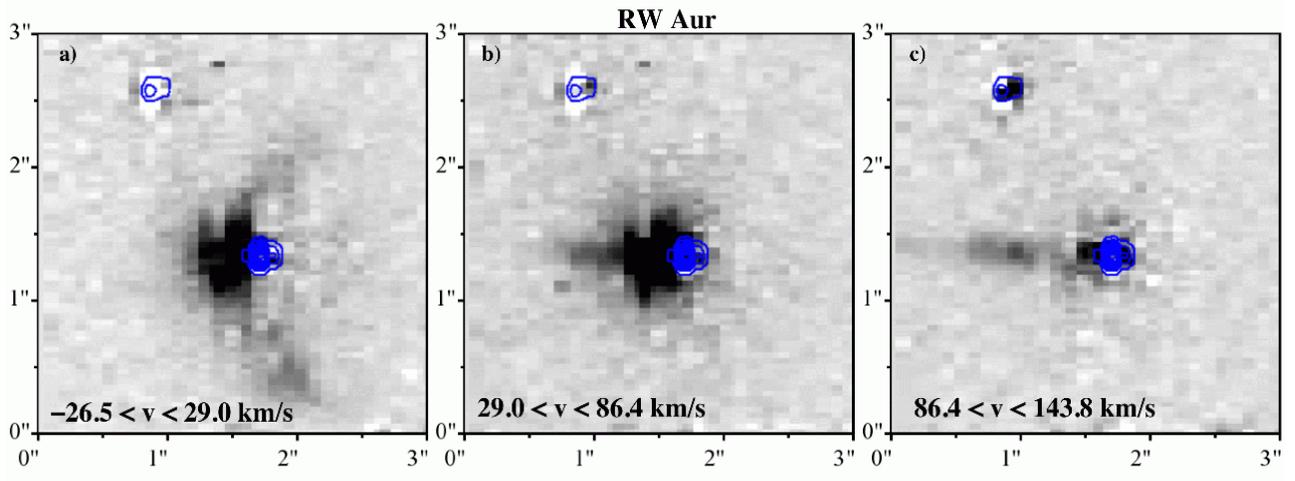}
\caption{Velocity channel map of spatially extended H$_2$ emission in the vicinity of RW Aur.  The three channel maps show H$_2$ from -26.5km/s to +143.8km/s where each panel is an average of two velocity slices through the H$_2$ datacube.}
\label{fig12}
\end{figure}

\begin{figure}
\centering
\includegraphics[angle=90, width=.45\textwidth]{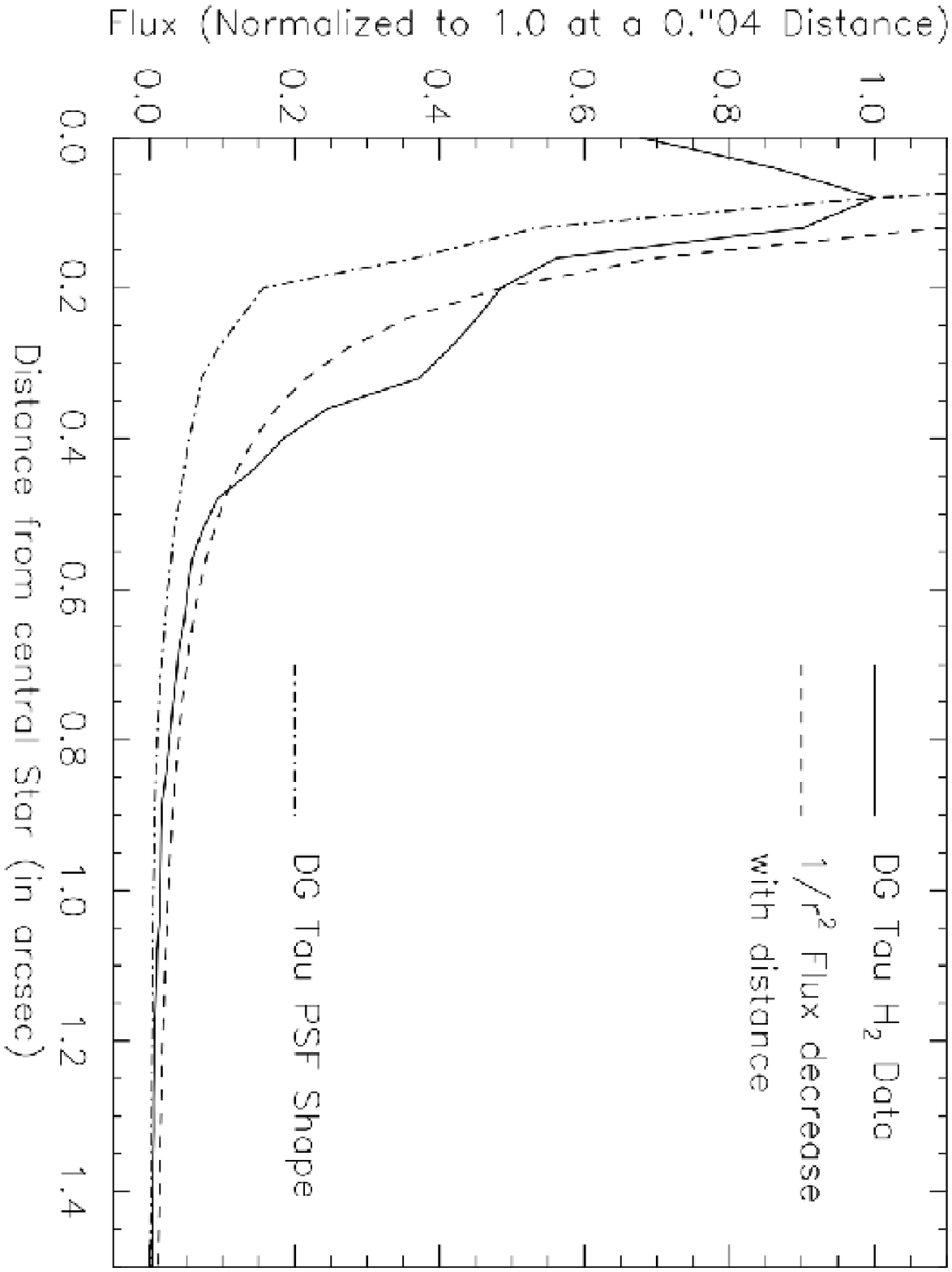}\ 
\includegraphics[angle=90, width=.45\textwidth]{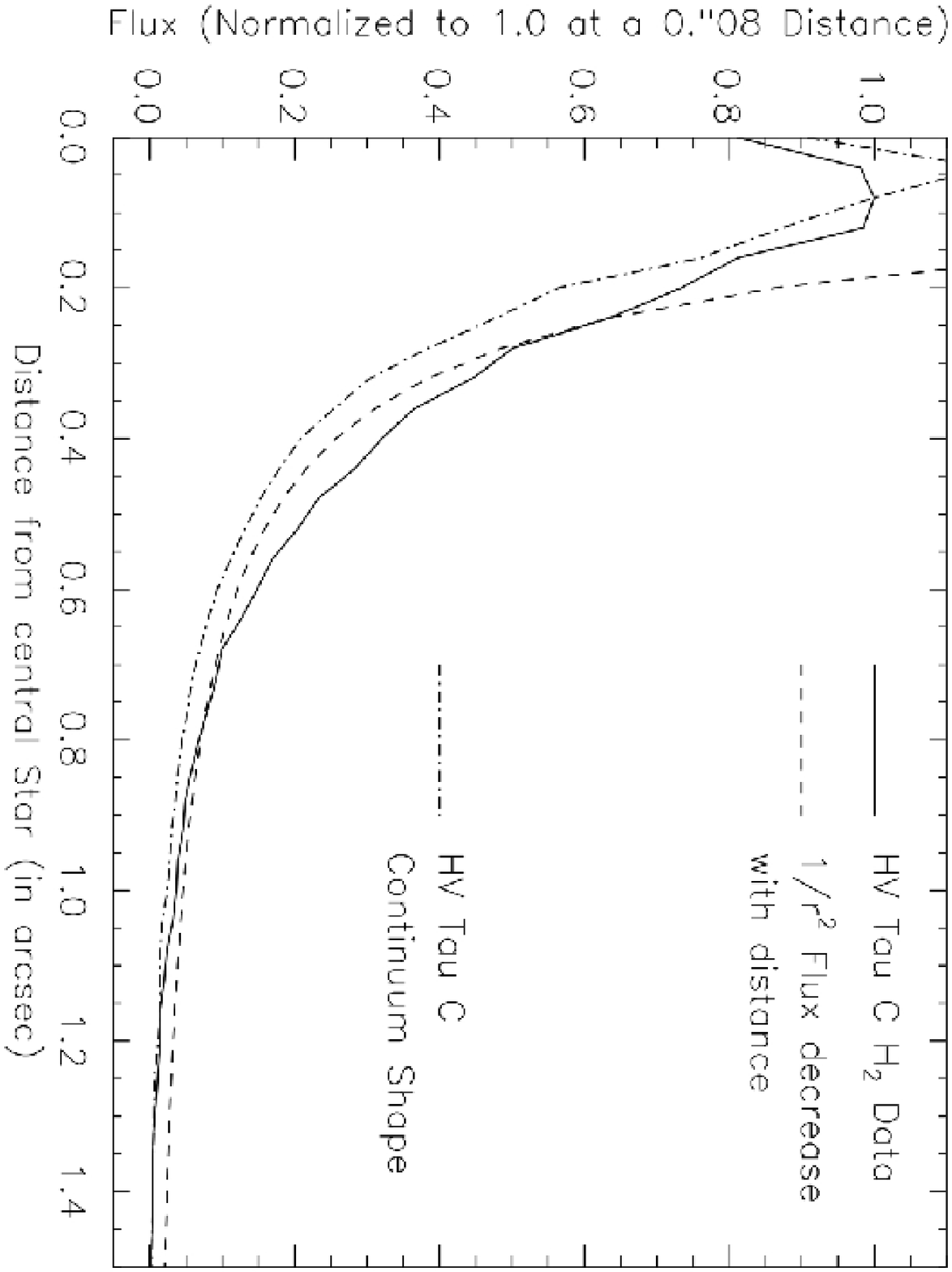}
\caption{Plots of the continuum subtracted H$_2$ flux as a function of increasing distance from the central star for DG Tau (a) and HV Tau C (b).  Overplotted on each is the shape of the continuum emission (dashed line) and, for comparison, flux that decreases with distance following a 1/r$^2$ power law (scaled to the H$_2$ flux at 0.$''$24). The H$_2$ and continuum fluxes were scaled to a normalization of 1.0 at a 0.$''$1 distance from the central star. The H$_2$ flux decreases at the central position of DG Tau results from 1 pixel of imperfect subtraction at the location of the brightest continuum flux, and for HV Tau C the H$_2$ flux decrease at the central position is a result of the dark lane seen in the edge-on disk geometry of this system (Figure 2).\label{fig13}}
\end{figure}

\end{document}